\documentclass[twocolumn]{aastex701}

\usepackage{bm}
\usepackage{mathtools}
\usepackage{mathrsfs}
\usepackage{tabularx}
\usepackage{algorithmic}
\usepackage{wrapfig}
\usepackage{float}
\usepackage{underscore}
\usepackage{xcolor}
\usepackage[capitalize,nameinlink]{cleveref}
\crefname{figure}{Fig.}{Figs.}
\Crefname{figure}{Figure}{Figures}

\begin{document}

\title{Cosmic Ray Reacceleration and Gamma-Ray Emission via the Cloud--Wind Interaction in the Galactic Halo}

\author[orcid=0009-0006-3783-2283,gname=Ryohei,sname=Inamoto]{Ryohei Inamoto}
\affiliation{Institute for Cosmic Ray Research, University of Tokyo, 5-1-5 Kashiwanoha, Kashiwa, Chiba 277-8582, Japan}
\email[show]{inamo714@icrr.u-tokyo.ac.jp}  

\author[orcid=0000-0003-3383-2279,gname=Jiro,sname=Shimoda]{Jiro Shimoda} 
\affiliation{Institute for Cosmic Ray Research, University of Tokyo, 5-1-5 Kashiwanoha, Kashiwa, Chiba 277-8582, Japan}
\email{jshimoda@icrr.u-tokyo.ac.jp}

\author[orcid=0000-0001-9064-160X,gname=Katsuaki,sname=Asano]{Katsuaki Asano}
\affiliation{Institute for Cosmic Ray Research, University of Tokyo, 5-1-5 Kashiwanoha, Kashiwa, Chiba 277-8582, Japan}
\email{asanok@icrr.u-tokyo.ac.jp}

\begin{abstract}
The interaction between cosmic rays (CRs) and multiphase gas
in the Galactic halo may play an important role in shaping
the hard gamma-ray spectrum of the Fermi bubbles.
We investigate the interaction between a hot Galactic wind
and a cold, dense gas clump with a relative velocity of $300\ {\rm km\,s^{-1}}$ using two-dimensional hydrodynamic simulations coupled with
momentum-dependent CR transport.
The system produces a bow shock and wind reacceleration region downstream, where the wind flow is faster than the initial velocity.
For a CR pressure comparable to the gas pressure (CR-rich), the wind acceleration and turbulence behind the clump are enhanced.
When CR pressure is dynamically weak, strong gas
compression leads to significant CR spectral hardening.
In contrast, for the CR-rich case, CR pressure gradients weaken
the gas compression and suppress further CR reacceleration.
We estimate the gamma-ray intensity from CRs, and find that
reaccelerated CRs penetrate the dense
clump and enhance hadronic gamma-ray emission.
The gamma-ray intensity reaches
$\sim 10^{-8}\,\mathrm{erg\,cm^{-2}\,s^{-1}\,sr^{-1}}$
in the 1--10 GeV band, and the gamma-ray spectra become significantly harder, consistent with the Fermi bubbles' emission.
These results suggest that an ensemble of dense
gas clumps interacting with the Galactic wind
may contribute to the gamma-ray emission of the Fermi bubbles.
\end{abstract}



\section{Introduction}

%Cosmic rays (CRs) are high-energy charged particles that constitute a nonthermal component of the interstellar medium (ISM). Although CRs have been extensively studied for more than a century, their acceleration sites and transport processes remain incompletely understood. In particular, the mechanisms that shape the nonthermal energy spectra observed near Earth are still under investigation.
Space telescopes performing direct detections of cosmic rays (CRs) such as AMS-02, CALET, and DAMPE provide precise CR spectra. However, only from those data, we cannot judge whether the CR spectra at the solar neighborhood are representative ones in our galaxy. The complex spectral breaks \citep{aguilar15,2019SciA....5.3793A,2022PhRvL.129j1102A} may suggest multiple types of acceleration sources \citep{2025NSRev..12af496L}, or they may indicate complex propagation processes spanning the galactic disk and halo \citep[e.g.,][]{2026ApJ..1003..114F}.
Turbulence in the interstellar medium (ISM), Galactic wind, and fresh gas accretion onto the Galactic halo should affect the CR propagation and reacceleration.
Gamma-ray observations provide an important means of probing CR populations beyond the solar neighborhood.

%High-energy gamma rays are produced through interactions between CRs and ambient matter or radiation fields, including neutral pion decay following inelastic proton-proton collisions and inverse Compton scattering by relativistic electrons.
%These emission processes allow us to investigate the spatial distribution and energy spectra of CRs throughout the Galaxy.

%One of the most remarkable discoveries in Galactic gamma-ray astronomy is t
The Fermi bubbles, a pair of giant gamma-ray structures extending approximately 10 kpc above and below the Galactic center \citep{Su2010,2012ApJ...753...61S,Ackermann2014}, may indicate the dynamical interaction between CRs and the halo gas.
The bubbles exhibit a relatively hard gamma-ray spectrum. %and a distinctive bipolar morphology, suggesting a connection to energetic processes in the central region of the Milky Way. However, both their physical origin and dominant gamma-ray emission mechanism remain uncertain.
While many models with explosive activities in the Galactic center, such as AGN jets \citep[e.g.,][]{Guo2012} and starbursts \citep[e.g.,][]{2014MNRAS.444L..39L}, have been proposed, CRs produced by long-term star formation can also be the source of the Fermi bubbles \citep[e.g.,][]{Crocker2011,ShimodaAsano2024}.

However, the origin of the hard gamma-ray spectrum of the Fermi bubbles remains an important question for hadronic scenarios.
The CR proton spectrum observed in the solar neighborhood \citep{aguilar15,2022PhRvL.129j1102A} is softer than that typically inferred from hadronic interpretations of the Fermi bubble emission \citep{Ackermann2014}.
Simple diffusion process of CRs cannot reconcile the observed spectra \citep{ShimodaAsano2024}.
This discrepancy suggests that CR reacceleration within the Galactic halo may play an important role in producing the hard gamma-ray spectrum of the Fermi bubbles.

In this context, the multiphase nature of the Galactic halo can provide a natural environment for CR reacceleration.
The Galactic halo is expected to contain both inflowing and outflowing gas as a consequence of gas accretion and Galactic winds \citep{Putman2012,shimoda2024}.
Observations of high-velocity clouds provide evidence for gas accretion onto the Galactic disk \citep{2007ApJ...670L.113W,2008ApJ...679L..21L}.
Furthermore, cold gas is also observed in the Galactic nuclear wind \citep{DiTeodoro2020,Lockman2026}, indicating the coexistence of multiple gas phases in the outflow. 
Motivated by these findings, several simulations have been performed in the context of gas accretion \citep[e.g.,][]{2026MNRAS.551g1405S} and galactic fountain \citep[e.g.,][]{2015MNRAS.447L..70F}.

In the model of \citet{ShimodaAsano2024}, the Fermi bubbles are produced via the interaction between CRs steadily supplied from the Galactic center and gases falling onto the Galactic disk.
In this model, the accompanying hot galactic wind, which is heated by CRs, is consistent with the eROSITA bubbles \citep{Predehl2020},
%Recent X-ray observations have provided additional constraints on the origin of the Fermi bubbles.
%\citet{Predehl2020} discovered the eROSITA bubbles, 
giant structures of diffuse thermal X-ray emission surrounding the Fermi bubbles.
%X-ray observations around the Fermi bubble edges indicate a characteristic plasma temperature of $k_{\rm B}T \sim 0.3$ keV \citep{Kataoka2013}, while the extended X-ray structures have a luminosity of order $L_{\rm X} \sim 10^{39}\ {\rm erg\,s^{-1}}$. These observations suggest that large-scale thermal gas structures are associated with energetic processes in the Galactic center.
Several studies point out that CRs should play an important role in the dynamics of Galactic winds \citep[e.g.,][]{Breitschwerdt1991, ShimodaInutsuka2022}.

Thus, the interaction between the hot Galactic wind, accreting cold gas, and CRs above the Galactic center is an important research target in the context of both Galaxy evolution and high-energy astrophysics.
Recent numerical studies have demonstrated that CRs can substantially modify the multiphase structure of Galactic winds and the evolution of cold gas clouds \citep{Armillotta2024, Thomas2025, Weber2025, Roy2026}.
Such wind--clump interactions can generate shocks and compressive flows, providing potential sites for CR reacceleration.
Previous studies have demonstrated that CR feedback can modify shock structures and drive hydrodynamic instabilities \citep{Drury1981, Drury1986, Haggerty2020}.
Recent galaxy-scale simulations have also begun to follow the time evolution of CR energy spectra together with gas dynamics \citep{Girichidis2022, Girichidis2024}.
However, how CR pressure modifies local reacceleration through its dynamical feedback during wind--clump interactions remains poorly understood.
To investigate this interplay, it is necessary to follow momentum-dependent CR transport together with gas hydrodynamics.

In this study, we investigate the interaction between a hot Galactic wind and a cold, dense clump using two-dimensional axisymmetric simulations that couple momentum-dependent CR transport with gas hydrodynamics.
By comparing models with different CR pressures, we examine how CR pressure affects the shock structure, gas compression, and flow dynamics around the clump.
We simultaneously follow the evolution of the CR spatial and energy distributions through spatial diffusion and momentum-space transport associated with gas compression and expansion, including the dynamical feedback of CR pressure on the gas.
Using the resulting CR and gas distributions, we calculate the spatial distribution and energy spectra of hadronic gamma-ray emission and investigate how CR feedback affects the emission intensity and morphology.
Through these analyses, we explore the role of CR--gas interactions in a multiphase Galactic wind and their possible contribution to the hard gamma-ray spectrum observed in the Fermi bubbles.

This paper is organized as follows. The simulation method to follow the interaction of hot gas and a cold dense cloud with CRs is presented in Section \ref{sec:method}.
The effects of different CR pressures on the gas dynamics, CR spectra, and gamma-ray emission are presented in Section \ref{sec:result}. 
Section \ref{sec:discussion} is devoted to discussion, especially for the Fermi bubble. Our conclusions are summarized in Section \ref{sec:summary}.

\section{Simulation Method}
\label{sec:method}

In this paper, we consider a dense gas clump accreting within the galactic wind and consistently solve the interaction between the gas clump, wind, and CRs.

\subsection{Basic equations}

We numerically solve the hydrodynamic equations coupled to CR transport
in the two-dimensional axisymmetric cylindrical coordinate $(R,z)$.
Adopting the mass density $\rho$, pressure $P$, and velocities $(v_R,v_z)$ of the fluid component representing the gas clump and the galactic wind,
the conservation equations for gas mass, momentum, and energy are given by
\begin{equation}
\frac{\partial \rho}{\partial t}
+
\frac{1}{R}\frac{\partial}{\partial R}
\left(R\rho v_R\right)
+
\frac{\partial}{\partial z}
\left(\rho v_z\right)
=0,
\end{equation}
\begin{equation}
\frac{\partial (\rho v_R)}{\partial t}
+
\frac{1}{R}\frac{\partial}{\partial R}
\left(R\rho v_R^2\right)
+
\frac{\partial}{\partial z}
\left(\rho v_R v_z\right)
=-\frac{\partial P}{\partial R}-\frac{\partial P_{\rm CR}}{\partial R},
\end{equation}
\begin{equation}
\frac{\partial (\rho v_z)}{\partial t}
+
\frac{1}{R}\frac{\partial}{\partial R}
\left(R\rho v_Rv_z\right)
+
\frac{\partial}{\partial z}
\left(\rho v_z^2+P\right)
=
-\frac{\partial P_{\rm CR}}{\partial z},
\end{equation}
and
\begin{equation}
\begin{aligned}
\frac{\partial e}{\partial t}
&+
\frac{1}{R}\frac{\partial}{\partial R}
\left[
R\left((e+P)v_R+q_R\right)
\right]
+
\frac{\partial}{\partial z}
\left[
(e+P)v_z+q_z
\right]
\\
&=
-v_R\frac{\partial P_{\rm CR}}{\partial R}
-v_z\frac{\partial P_{\rm CR}}{\partial z}
+Q(n,T),
\label{eq:ene}
\end{aligned}
\end{equation}
respectively. Here, we have neglected wave excitation by CRs, which appears in CR--MHD systems and may heat the background fluid eventually. As can be understood from the energy equation, the CR pressure $P_{\rm CR}$ acts as an external force on the fluid, so that CRs do not directly affect the entropy of the background fluid. The total gas energy density is written as
\begin{equation}
e=
\frac{P}{\gamma-1}
+
\frac{1}{2}\rho
\left(v_R^2+v_z^2\right),
\label{eq:gas-ene-den}
\end{equation}
%Here, $\rho$ is the gas mass density, $n$ is the gas number density, $\mbox{\boldmath $E$}{v}=(v_R,v_z)$ is the gas velocity, $P$ is the thermal pressure, and $e$ is the total gas energy density.
with an adiabatic index $\gamma=5/3$.

The last term $Q=-n\mathcal{L}(n,T)$ in the energy equation represents radiative energy change, which is written with a function
\begin{equation}
\mathcal{L}(n,T)=n\Lambda(T)-\Gamma_0,
\end{equation}
where the first and second terms are radiative cooling and heating terms, respectively. For the cooling function $\Lambda(T)$, we adopt the formula proposed in \citet{Kakiuchi2024ApJ}\footnote{In this formula, the formulae for $T<10^4$ K \citep{2002ApJkoyamainutsuka...564L..97K} and for $T>10^4$ K \citep{SutherlandDopita1993} are smoothly connected.},
where the density $n$ and temperature $T$ are defined as $n=\rho/(\mu m_{\rm p})$ and $T=P/n$.
% are related by $\rho=\mu m_p n$, where $m_p$ is the proton mass.
For simplicity, we adopt a constant mean molecular weight of $\mu=0.6$,
corresponding to the fully ionized limit.
{In most parts of our computational domain where the flow is dynamically turbulent, the temperature is well above $10^4$ K, so the difference in the fluid motion resulting from the constant $\mu$ is likely to be small.}

%The last term in the gas energy equation represents radiative cooling and heating, which we write as
%
%\begin{equation}
%\mathcal{L}(n,T)=n\Lambda(T)-\Gamma_0,
%\end{equation}
%
%such that the corresponding %volumetric loss rate is
%$n\mathcal{L}=n^2\Lambda(T)-n\Gamma_0$.
%For the low-temperature cooling function, we adopt the fitting formula of \citet{2002ApJkoyamainutsuka...564L..97K}, using the corrected expression, 
%
%\begin{equation}
%\begin{aligned}
%\Lambda_{\rm low}(T)
%=2.0\times10^{-26}
%\Bigg[
%&10^7\exp\left(-\frac{118400}{T+1000}\right)\\
%&+1.4\times10^{-2}T^{1/2}
%\exp\left(-\frac{92}{T}\right)
%\Bigg],
%\end{aligned}
%\end{equation}
%
%in units of $\mathrm{erg\,cm^3\,s^{-1}}$. This is smoothly connected to the high-temperature cooling function of \citet{SutherlandDopita1993} around $T\sim10^4$ K, following \citet{Kakiuchi2024ApJ}.
For the heating term, we assume photoelectric heating by dust grains and adopt a constant
heating rate of $\Gamma_0=2.0\times10^{-25}\ \mathrm{erg\,s^{-1}}$,
motivated by the enhanced interstellar radiation field in the Galactic
center environment. Since we focus on a local region of the Galactic
halo, we neglect spatial variations in the heating rate. For simplicity,
we also neglect its temperature dependence.

Thermal conduction is included through the conductive energy flux.
The standard heat flux is defined as
\begin{equation}
\mbox{\boldmath $q$}_0
=
-\kappa(T) \left( \frac{\partial T}{\partial R} , \frac{\partial T}{\partial z} \right),
\end{equation}
where $\kappa(T)$ is the thermal conductivity.
We adopt the thermal conductivities
$\kappa_{\rm high}=1.24\times10^{-6}T^{5/2}$
for ionized gas \citep{Spitzer1962} and
$\kappa_{\rm low}=2.5\times10^3 T^{1/2}$
for neutral gas \citep{1953ApJparker...117..431P},
in cgs units.
To smoothly connect the low- and high-temperature regimes, we use
\begin{equation}
\kappa(T)
=
\left[
\kappa_{\rm low}^2(T)
+
\kappa_{\rm high}^2(T)
\right]^{1/2}.
\end{equation}
In a hot and rarefied plasma, the electron mean free path can become
comparable to the characteristic temperature-gradient scale,
invalidating the classical diffusion approximation.
We therefore introduce a saturated heat flux following
\citet{Cowie1977},
$
q_{\rm sat}
=
5\phi_{\rm sat}\rho c_s^3 ,
$
where $\phi_{\rm sat}$ is a dimensionless coefficient and
$c_s=\sqrt{\gamma P/\rho}$ is the gas sound speed.
We adopt $\phi_{\rm sat}=0.3$, appropriate for a fully ionized gas
\citep{1984ApJgiuliani...277..605G,2005A&Aorlando...444..505O}.
The conductive flux in equation (\ref{eq:ene}) is then written as
\begin{equation}
\mbox{\boldmath $q$}
=(q_R, q_z)=
\frac{\mbox{\boldmath $q$}_0}
{1+|\mbox{\boldmath $q$}_0|/q_{\rm sat}} .
\end{equation}

The time evolution of the CR momentum distribution $\psi_{\rm CR}(R,z,p)$ is regulated by the motion of the background fluid. Assuming isotropic momentum distributions, we numerically solve the diffusion--convection equation
\begin{eqnarray}
\frac{\partial\psi_{\rm CR}}{\partial t}
&=&
\nabla\cdot
\left[
D(p)\nabla\psi_{\rm CR}
-
\mbox{\boldmath $v$}\psi_{\rm CR}
\right]
+
\frac{\partial}{\partial p}
\left[
\frac{p}{3}
(\nabla\cdot\mbox{\boldmath $v$})
\psi_{\rm CR}
\right]\nonumber \\
&=&
\frac{1}{R}\frac{\partial}{\partial R}
\left(
R D(p) \frac{\partial}{\partial R} \psi_{\rm CR}
-
R v_R \psi_{\rm CR}
\right)
\nonumber \\
&+&
\frac{\partial}{\partial z}
\left(
D(p) \frac{\partial}{\partial z} \psi_{\rm CR}
-
v_z \psi_{\rm CR}
\right)
\nonumber \\
&+&
\frac{\partial}{\partial p}
\left[
\frac{p}{3}
\left( \frac{1}{R}\frac{\partial R v_R}{\partial R} + \frac{\partial v_z}{\partial z} \right)
\psi_{\rm CR}
\right],
\end{eqnarray}
where $D(p)$ is the spatial diffusion coefficient.
For simplicity, we assume here that the diffusion coefficient is spatially uniform, regardless of the properties of the background gas.
Motivated by
Kolmogorov turbulence, we adopt
\begin{equation}
D(p)
=
D_0
\left(
\frac{p}{1\ {\rm GeV}/c}
\right)^{1/3},
\end{equation}
with
$
D_0=1.0\times10^{27}\ {\rm cm^2\,s^{-1}} .
$
%
%This coefficient is chosen to be one order of magnitude lower than the canonical Galactic diffusion coefficient
%to reduce the computational cost. However, around the turbulent regions, such a locally suppressed value is possible \citep{hawc17}.
This coefficient is chosen to be one order of magnitude lower than the canonical Galactic diffusion coefficient to reduce the computational cost. However, considering relatively strong magnetic fields inferred in the Fermi bubble region ($B\sim 10\,\mu{\rm G}$; \citealt{Carretti2013, Ackermann2014}), such a locally suppressed diffusion coefficient is plausible in turbulent regions \citep{hawc17}.

%in the solar neighborhood  and represents suppressed CR transport in the environment considered here.

The CR pressure is calculated %from the isotropic CR momentum distribution 
as
\begin{equation}
P_{\rm CR}
=
\frac{1}{3}
\int dp\,
p\,v(p)\,\psi_{\rm CR}(p),
\end{equation}
where
\begin{equation}
v(p)
=
\frac{pc^2}
{\sqrt{(pc)^2+(m_{\rm p} c^2)^2}}
\end{equation}
is the particle velocity.

\subsection{Numerical Procedure}

The hydrodynamic equations are solved using the HLL approximate Riemann solver \citep{Harten1983}.
We adopt the second-order MUSCL scheme \citep{1979JCoPh..32..101V} with a minmod slope limiter for spatial reconstruction and the second-order Runge--Kutta scheme for time integration.
For the MUSCL reconstruction,  we choose the logarithms of the gas density and entropy, along with the linear velocity components, as the reconstruction variables. This choice accounts for the adiabatic relation while allowing the reconstruction to handle large density contrasts and maintain positive density and pressure at cell interfaces.
The spatial reconstruction is locally reduced to first order in regions with large density contrasts and expanding flows ($\nabla\cdot\mbox{\boldmath $v$}>0$) to maintain numerical stability.

The CR transport equation is solved using an upwind scheme for the advection terms and an explicit scheme for spatial diffusion.

\subsection{Gamma-Ray Emissivity}

Based on the simulation outputs for gas and CRs, we calculate the hadronic gamma-ray emission following the method
of \citet{Nishiwaki2021}.
Following \citet{Kelner2006}, the gamma-ray emissivity
from neutral-pion decay is evaluated as
\begin{equation}
    q_\gamma(E_\gamma,\boldsymbol{x})
    =
    2 E_\gamma \int_{E_{\pi,{\rm min}}}^{\infty}
    \frac{
        q_{\pi^0}(E_\pi,\boldsymbol{x})
    }{
        \sqrt{E_\pi^2-m_\pi^2c^4}
    }
    \,dE_\pi ,
    \label{eq:qgamma}
\end{equation}
where $E_\pi$ is the neutral-pion energy, and
\begin{equation}
    E_{\pi,{\rm min}}
    =
    E_\gamma+\frac{m_\pi^2c^4}{4E_\gamma}.
\end{equation}
Given $E_\pi$, the gamma-ray photons distribute in the energy range of $E_\pi (1-\beta_\pi)/2 \leq E_\gamma \leq E_\pi (1+\beta_\pi)/2$ with $\beta_\pi=\sqrt{1-m_\pi^2 c^4/E_\pi^2}$.
The injection rate of neutral pions %emissivity
is calculated from
the local gas density and CR proton spectrum as
\begin{equation}
    q_{\pi^0}(E_\pi,R,z)
    =
    n_{\rm gas}
    \int dp\,
    \psi_{\rm CR}(p)\,v(p)\,
    \frac{d\sigma_{pp\rightarrow\pi^0}(E_{\rm p},E_\pi)}
    {dE_\pi},
    \label{eq:qpi}
\end{equation}
where $n_{\rm gas} =\rho/m_{\rm p}$ is the target gas density,
$\psi_{\rm CR}$ is the CR proton number density
per unit momentum, and $v_p$ is the proton velocity.
The CR proton %total and kinetic energies
energy is given by
\begin{equation}
    E_{\rm p}=\sqrt{p^2c^2+m_{\rm p}^2c^4}.
%    \qquad
%    T_p=E_p-m_pc^2.
\end{equation}
For the differential cross section of neutral-pion
production $d\sigma_{pp\rightarrow\pi^0}/dE_\pi$, we adopt the formulae of
\citet{Kamae2006,Kamae2007}.

%% outer boundaryの話も書いておく
%% initial conditionの話; line cooling, initial condition
%% \Gamma_0 = 1.0×10^-25 erg/s, v_w=300 km/sの結果で議論する
%%piecewise_constantの説明をもう少し

%%%%%%%%%%%%%%%%%%%%%%%%%%%%%%%
%%%%%%%%%%%%%%%%%%%%%%%%%%%%%%%
%%%%%%%%%%%%%%%%%%%%%%%%%%%%%%%
%%%%%%%% Result %%%%%%%%%^^^^^^^^^^^^^^^^^^^^^^^^^^^^^
%%%%%%%%%%%%%%%%%%%%%%%%%%%%%%%
%%%%%%%%%%%%%%%%%%%%%%%%%%%%%%%
%%%%%%%%%%%%%%%%%%%%%%%%%%%%%%%

\section{Simulation Results}
\label{sec:result}

\subsection{Simulation Setup}

We consider a local region of the Galactic halo in which a cold gas
clump interacts with an upward hot Galactic wind.
The computational domain covers $0\leq R\leq450$ pc and
$0\leq z\leq800$ pc, with $225\times400$ spatial cells
and a uniform cell size of 2 pc.
The CR distribution is simultaneously evolved over the momentum range
$0.1\leq p\leq1000$ GeV/$c$ using 24 momentum bins.

Initially, the cold dense clump with a radius of 50 pc is embedded in a hot rarefied ambient medium.
While the gas is divided into two phases, the cold clump and hot wind, the pressure is spatially uniform as $P=1.0\ {\rm eV\,cm^{-3}}$, and heating and cooling are in equilibrium as $\mathcal{L}(n,T)=0$.
To realize the above requirement, we adopt gas number densities and temperatures of the clump and wind as $
(n_{\rm c},T_{\rm c})
=
(1.75\ {\rm cm^{-3}},\,6.61\times10^3\ {\rm K})
$ and $
(n_{\rm w},T_{\rm w})
=
(4.72\times10^{-3}\ {\rm cm^{-3}},\,
2.46\times10^6\ {\rm K})
$, respectively.
%denoted by $(n_c,T_c)$ and $(n_w,T_w)$
The computation frame is the clump rest frame.
The injected wind velocity is $v_w=300\ \mathrm{km\,s^{-1}}$.
The velocity of the wind flowing from bottom to top corresponds to the relative velocity between the accreting clump and the Galactic wind.
%gas pressure is spatially uniform across the computational domain. The gas number densities and temperatures of the clump and wind are denoted by $(n_c,T_c)$ and $(n_w,T_w)$, respectively. Both phases are initially in thermal equilibrium,
%
%\begin{equation}
%\mathcal{L}(n_i,T_i)=0,
%\qquad
%i=c,w.
%\end{equation}
%

The initial CR energy density is also uniform, and its spectrum is assumed to follow a broken power law in
momentum as
\begin{equation}
\psi_{\rm CR}(p)
\propto
\begin{cases}
p^{-2.7}, & p>1\ {\rm GeV}/c,\\
p,        & p\leq1\ {\rm GeV}/c.
\end{cases}
\end{equation}
The high-momentum index $-2.7$ is intended to represent Galactic CRs
that have escaped from acceleration sites in the disk and have been
softened during propagation into the Galactic halo
\citep{Strong2007ARNPS}.

We compare two calculations that differ only in the initial CR pressure.
In the CR-rich model, the CR pressure is initially equal to the thermal
gas pressure,
\begin{equation}
P_{\rm CR,0}=P_0=1.0\ {\rm eV\,cm^{-3}},
\end{equation}
whereas in the CR-poor model, the initial CR pressure is 100 times
smaller,
\begin{equation}
P_{\rm CR,0}=0.01\,P_0=0.01\ {\rm eV\,cm^{-3}}.
\end{equation}
The latter therefore provides a limiting case in which CR feedback is
initially dynamically negligible.
%Throughout this paper, we adopt the rest frame of the clump.
At the symmetry axis, $R=0$, we impose the axisymmetric boundary
condition $v_R=0$. At the outer radial boundary, $R=450$ pc,
we impose an outflow boundary condition.
At the upper boundary, $z=800$ pc, we adopt an escape boundary for CRs.
At the lower boundary, $z=0$, the wind is continuously injected with
the same gas and CR properties as the initial ambient medium and with
an upward velocity $v_z=300\ {\rm km\,s^{-1}}$.

As we have mentioned, the heating rate and diffusion coefficient are uniform and constant with
$\Gamma_0=2.0\times10^{-25}\ \mathrm{erg\,s^{-1}}$ and
$
D_0=1.0\times10^{27}\ {\rm cm^2\,s^{-1}}
$, respectively.

\subsection{Gas Dynamics}

% ============================================================
% Figures 1--2: Two-dimensional gas evolution
% ============================================================
\begin{figure*}
    \centering
    \includegraphics[width=1.0\linewidth]{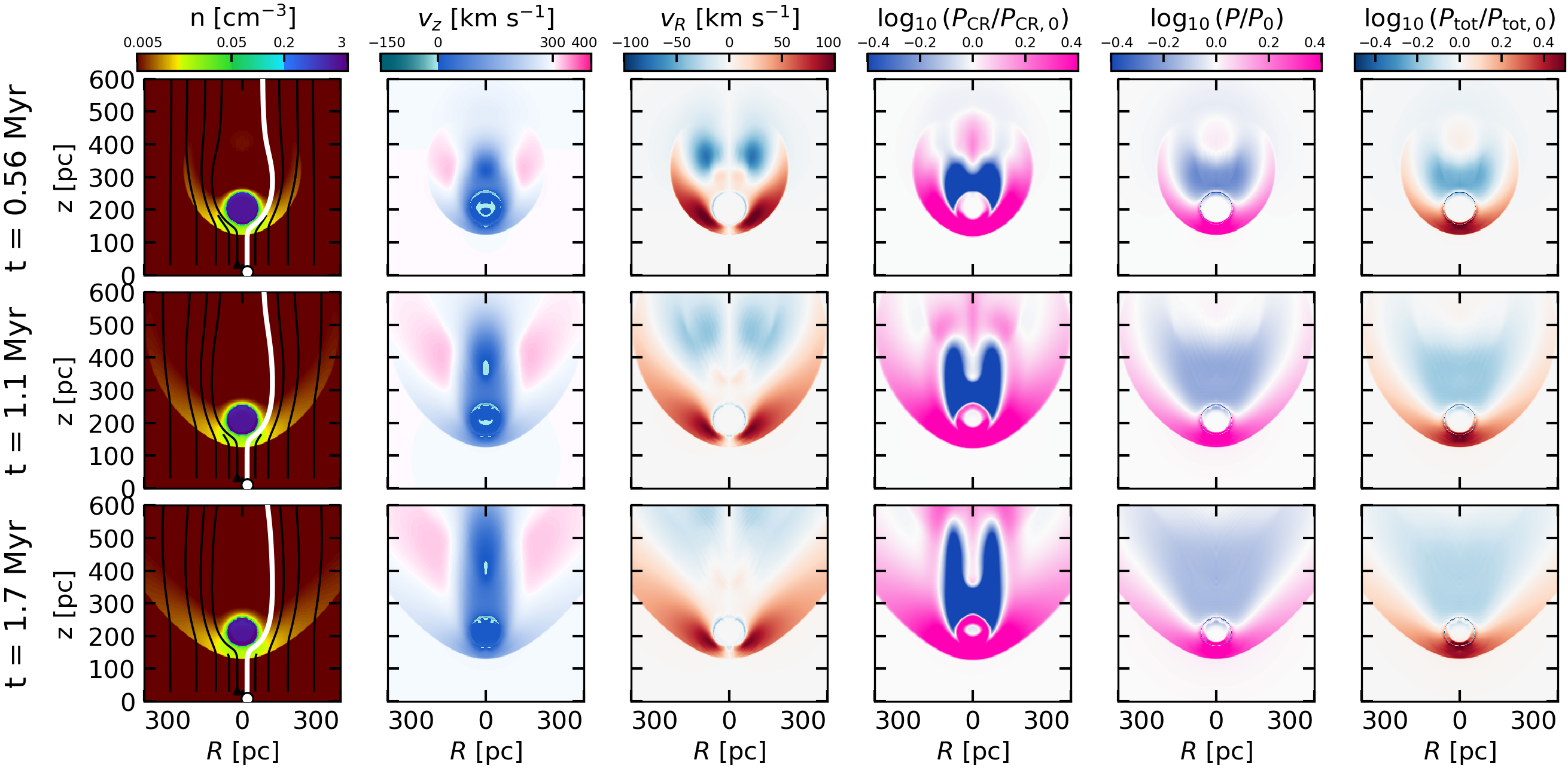}
    \caption{
    Time evolution of the clump--wind interaction in the CR-poor model.
    From left to right, the columns show the gas density $n$, vertical
    velocity $v_z$, radial velocity $v_R$, CR pressure $P_{\rm CR}$, gas pressure $P$, and total pressure $P_{\rm tot}=P+P_{\rm CR}$. The pressures are shown normalized to their initial values. %relative to its initial value, $\log_{10}(P_{\rm CR}/P_{{\rm CR},0})$, the gas pressure relative to its initial value, $\log_{10}(P/P_0)$, and the total pressure relative to its initial value, $\log_{10}(P_{\rm tot}/P_{{\rm tot},0})$, where $P_{\rm tot}=P+P_{\rm CR}$.
    The simulation is axisymmetric, and each panel is mirrored about the
    symmetry axis for visualization.
    Black curves in the density panels show representative streamlines,
    while the thick white curve indicates the streamline used for the
    one-dimensional analysis in Fig.~\ref{fig:1devolution_thin_cr}.
    }
    \label{fig:timeevolution_thin_cr}
\end{figure*}
\begin{figure*}
    \centering
    \includegraphics[width=1.0\linewidth]{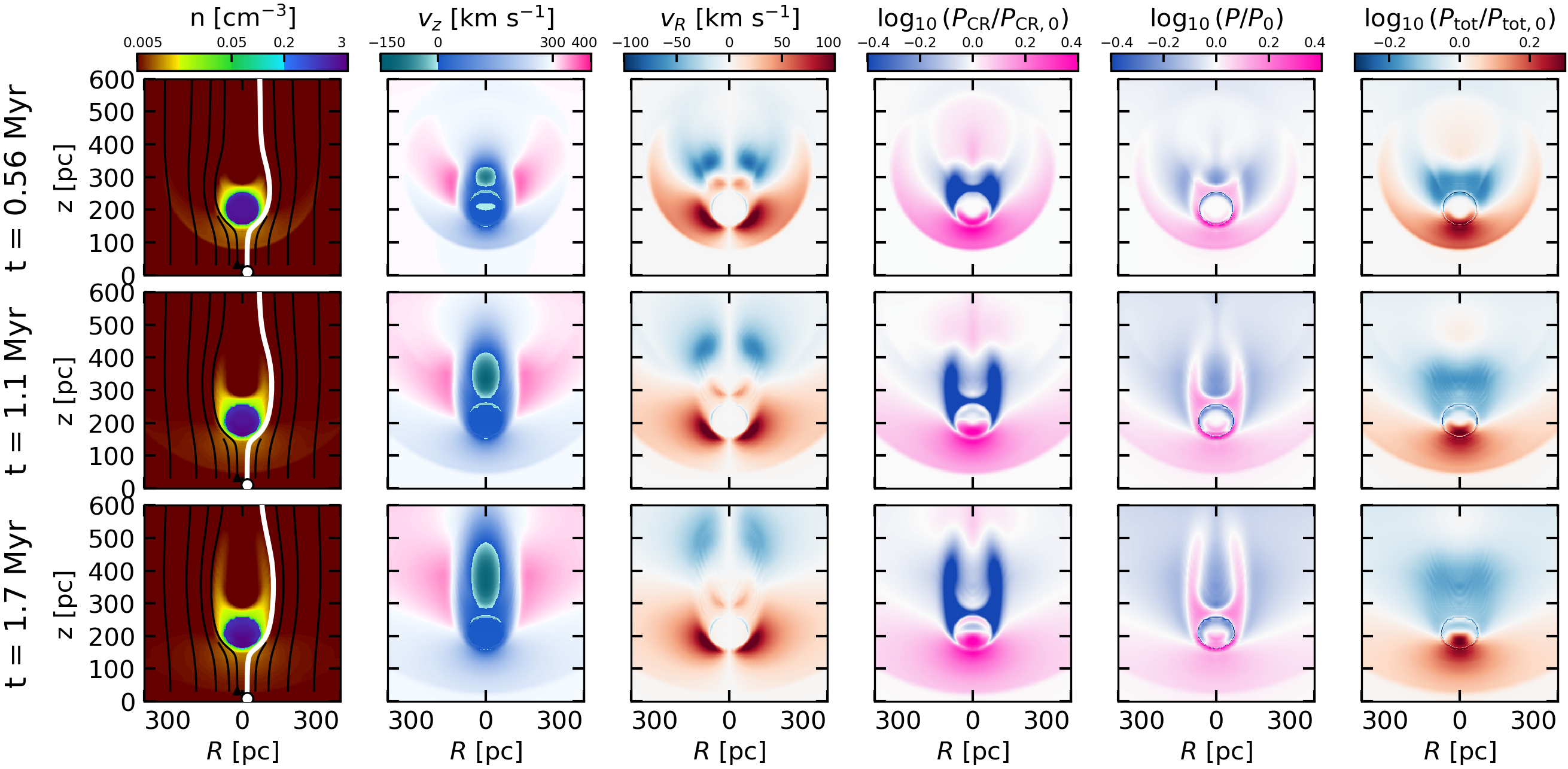}
    \caption{
    Same as Fig.~\ref{fig:timeevolution_thin_cr}, but for the CR-rich model. %    in which the initial CR pressure is equal to the thermal gas pressure,    $P_{{\rm CR},0}=P_0$.
    The thick white curve in the density panels indicates the streamline used
    for the one-dimensional analysis in Fig.~\ref{fig:1devolution_thick_cr}.
    }
    \label{fig:timeevolution_thick_cr}
\end{figure*}
In this section, we examine how CR pressure affects gas dynamics in the
interaction between the Galactic wind and the cold gas clump.
Figures~\ref{fig:timeevolution_thin_cr} and
\ref{fig:timeevolution_thick_cr} show the evolution of gas dynamics in the CR-poor and CR-rich cases, respectively.
In both models, the incoming wind encounters the cold,
dense clump and forms a bow shock.
As shown in the pressure maps, the curvature of the shock fronts grows with time.
In the CR-poor model, the bow shock is characterized by
a relatively sharp density jump. %and a localized region of strong gas compression.
In contrast, the CR-rich model exhibits a weaker density
jump and the distance from the shock front to the clump head is longer than that in the CR-poor model. %a broader compression region extending farther upstream from the clump. The {\bf volume of} the enhanced CR pressure also spreads over a larger region around the clump.

The two models also differ in their downstream structures.
As shown in the velocity maps ($v_z$ and $v_R$), the gas behind the clump is decelerated, while the outer gas is accelerated faster than its initial velocity.
%As the wind flows around the clump, a wake develops behind it, accompanied by regions of negative vertical velocity, $v_z<0$.
These regions grow more prominently in the CR-rich model,
where the downstream flow becomes increasingly disturbed
and develops more complex velocity structures.
The region of negative velocity ($v_z<0$) behind the clump significantly grows with time.

The CR pressure maps show horn-like structures behind the clump, where the CR pressure is low. This structure arises because the density of CRs originally present within the clump decreases owing to gas expansion driven by the separation flow from the rear of the clump, generating rarefaction waves. This is the result of the uniform CR distribution in the initial condition. In the CR-rich case, this structure affects the gas motion, resulting in a similar shape of a high gas-pressure region. This spatial anticorrelation between the gas and CR pressures
%spatial separation 
highlights the importance of solving CR transport in simulations %independently of the gas density, 
rather than imposing a prescribed relation between the two.

% ============================================================
% Figures 3--4: Streamline profiles and CR-modified shock
% ============================================================
\begin{figure*}
    \centering
    \includegraphics[width=1.0\linewidth]{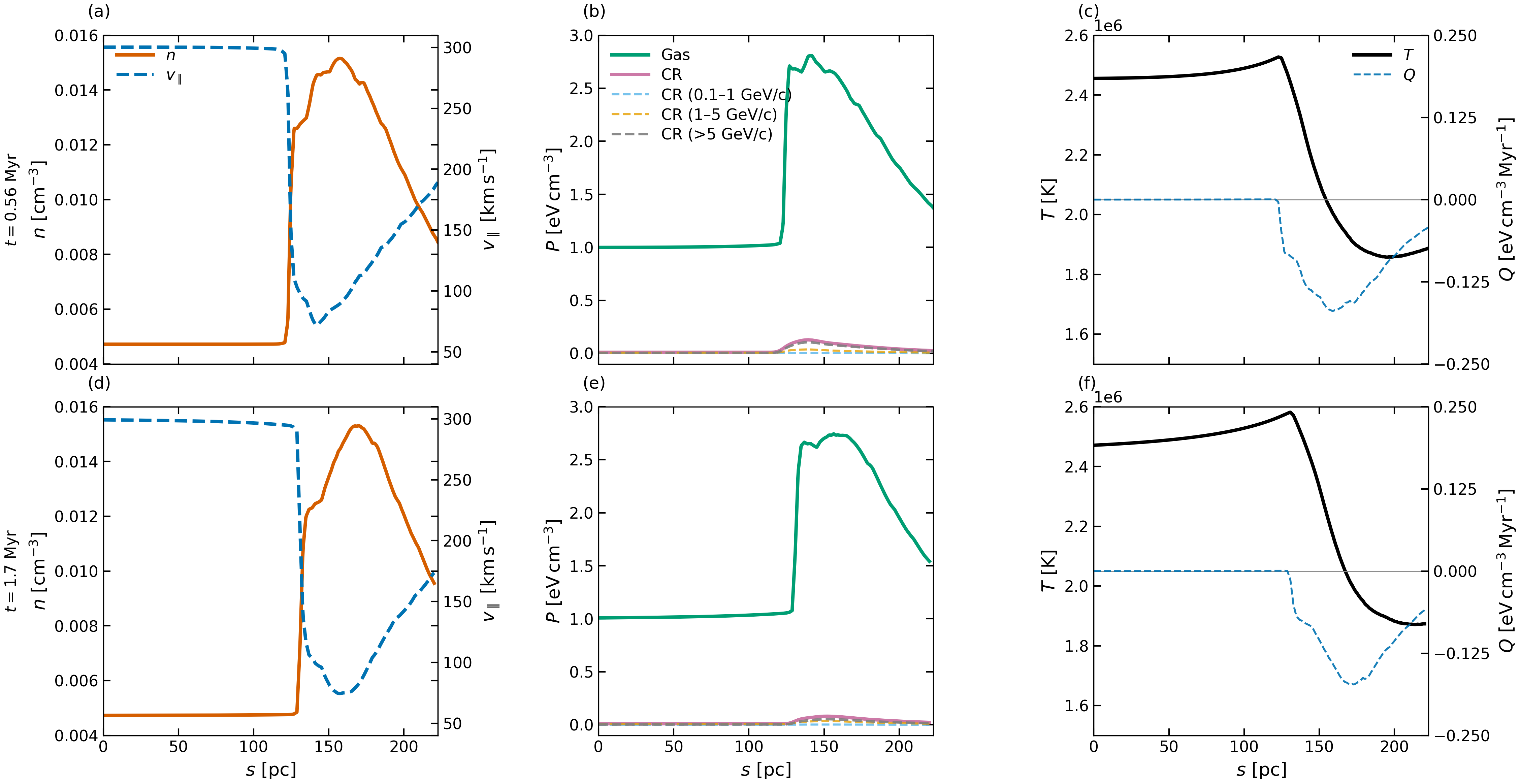}
    \caption{
    One-dimensional profiles along the representative streamline highlighted
    in Fig.~\ref{fig:timeevolution_thin_cr} for the CR-poor model at
    $t=0.565$ and $1.695$ Myr. The horizontal axes show the distance along the streamline $s$.
    The left panels show the gas density $n$ and the velocity component
    parallel to the streamline, $v_{\parallel}$.
    The middle panels show the thermal gas pressure $P_{\rm gas}$ and
    CR pressure $P_{\rm CR}$, while the right panels show the gas temperature
    $T$ and the net radiative heating rate $Q=-n\mathcal{L}(n,T)$,
    where negative values indicate net cooling.
    }
    \label{fig:1devolution_thin_cr}
\end{figure*}
\begin{figure*}
    \centering
    \includegraphics[width=1.0\linewidth]{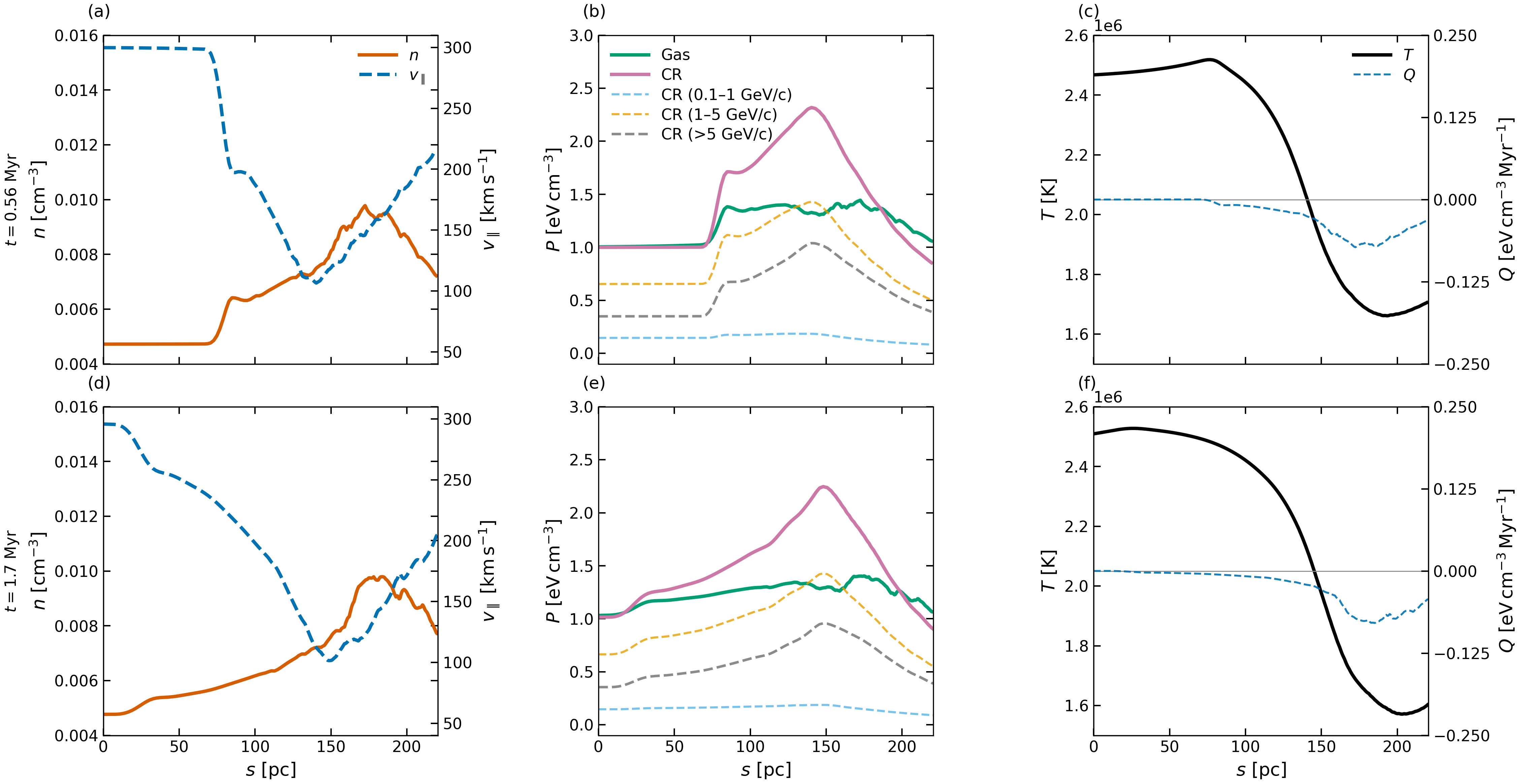}
    \caption{
    Same as Fig.~\ref{fig:1devolution_thin_cr}, but for the CR-rich model.
    }
    \label{fig:1devolution_thick_cr}
\end{figure*}

In Figures~\ref{fig:1devolution_thin_cr} and
\ref{fig:1devolution_thick_cr}, we also plot the gas and CR properties along the representative
streamlines shown in the density maps in Figures~\ref{fig:timeevolution_thin_cr} and
\ref{fig:timeevolution_thick_cr}.
%Figures~\ref{fig:1devolution_thin_cr} and \ref{fig:1devolution_thick_cr} present the corresponding one-dimensional profiles for the CR-poor and CR-richmodels, respectively. 
This streamline heads toward the clump from nearly directly in front. After passing through the shock wave, it flows toward the back of the clump, as if bypassing it. However, since this streamline is based on a snapshot, it does not necessarily represent the steady flow in general.

In the CR-poor model, a clear shock front is seen, and it has been maintained throughout the simulation. The density further increases behind the shock because the downstream flow is influenced by the clump and radiative cooling.
%In the CR-poor model, a slight spatial offset is observed between the sharp decrease in $v_\parallel$ and the increase in gas density, as shown in panels (a) and (d) of Fig.~\ref{fig:1devolution_thin_cr}.
As shown by the radial velocity distributions in
Fig.~\ref{fig:timeevolution_thin_cr}, the gas passing
through the bow shock is strongly deflected radially
outward ($v_R>0$) as it flows around the clump.
%The velocity decrease along the streamline reflects the passage through the bow shock, while the subsequent density evolution is influenced by the deflected downstream flow.
This two-dimensional flow geometry, radiative cooling (see temperature profile), and the unsteady flow may account for
 the offset between the velocity minimum and density peak.
%the slight delay in the density increase relative to the sharp velocity decrease.
Behind the density peak position, the gas expands and accelerates. As can be seen from the pressure distribution in panels (b) and (e) of Figure \ref{fig:1devolution_thin_cr}, the CR effect is almost negligible in the CR-poor case.

The shocked gas departs from thermal equilibrium and enters a state dominated by radiative cooling. As shown in panels (c) and (f) of
Fig.~\ref{fig:1devolution_thin_cr}, the gas evolves almost isothermally. A significant temperature jump at the shock does not appear.
%The compressed gas subsequently undergoes radiative cooling. In the CR-poor model, the net cooling rate increases in the high-density region downstream of the shock, 
The enhanced cooling is accompanied by a decrease
in gas temperature and further compression of
the post-shock gas.

%the incoming gas remains nearly unchanged until it reaches a narrow compression region, where the velocity decreases rapidly and the density and thermal pressure increase sharply.
In the CR-rich model, the gas flow begins to
decelerate well upstream of the density peak.
As shown in Figure \ref{fig:1devolution_thick_cr},
the CR pressure dominates the region behind the shock. The density jump is weak, and the CR pressure distribution gradually evolves into a spatially broader and smoother profile. The CR effect is also evident in the smooth temperature distribution.
%This gradual deceleration coincides with an increase in the CR pressure, which becomes comparable to or exceeds the thermal pressure.
This behavior is characteristic of a CR-modified shock \citep{Drury1981},
in which CRs diffusing upstream exert a pressure gradient
on the incoming gas and form a precursor ahead of the
main compression region.
The CR-pressure gradient gradually decelerates the
incoming wind before it reaches the main compression
front, thereby broadening the velocity transition and
weakening the gas compression across the front.
%Our results demonstrate that such CR modification of shock structures also occurs in the interaction between a Galactic halo wind and a cold gas clump.

\begin{figure}
    \centering
    \includegraphics[width=1.0\linewidth]{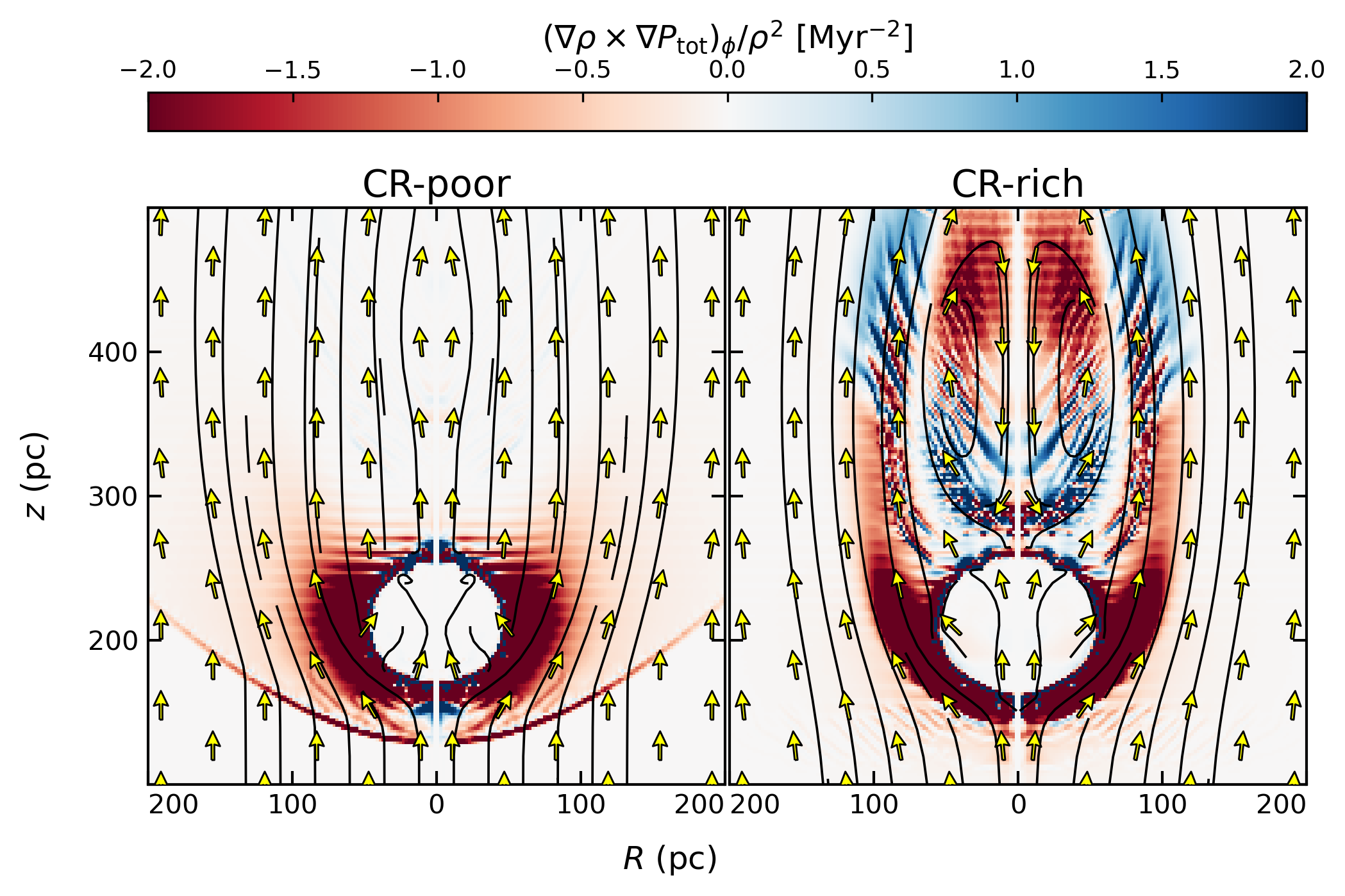}
    \caption{Baroclinic vorticity source term in the CR-poor (left) and CR-rich (right) models at 1.7~Myr. Black curves show gas streamlines, and yellow arrows indicate the local flow direction.}
    \label{fig:baroclinic}
\end{figure}

The horn-like structures {of the CR pressure behind the clump}
%in the downstream wake of the CR-rich model 
(Fig.\ref{fig:timeevolution_thick_cr}) induce the dynamical influence on the background gas. %provide another example of the dynamical influence of CR pressure. 
%As the wake develops, CRs are advected and redistributed by the circulating flow, leading to enhanced CR pressure near the symmetry axis. 
The resulting CR pressure gradient acts as an external force and pushes the surrounding gas. This CR effect causes differences in the development of turbulence in the CR-poor and CR-rich cases. As shown in Figure~\ref{fig:baroclinic}, the turbulence behind the clump is enhanced in the CR-rich case compared to the CR-poor case. A swirling flow is formed behind the clump in the CR-rich case.
%radially outward, while the fast Galactic wind continues to flow around the outer sides of the clump. The gas between these regions is consequently compressed into horn-like structures. Notably, the CR pressure inside the horns remains relatively low despite the enhanced gas density. 

%%%%%%%%%%%%%%%%%%%%%%%% stream line %%%%%%%%%%%%%%%%%%%%%%%%%%%
To clarify the generation mechanism of rotational gas motion, %during the clump--wind interaction.
we also show the baroclinic
source term in Figure~\ref{fig:baroclinic}.
The gas vorticity is defined as
$\boldsymbol{\omega}\equiv\nabla\times\mbox{\boldmath $v$}$.
Taking the curl of the momentum equation gives
\begin{equation}
\frac{\partial\boldsymbol{\omega}}{\partial t}
+\nabla\times(\boldsymbol{\omega}\times\boldsymbol{v})
=
\frac{\nabla\rho\times\nabla P_{\rm tot}}{\rho^2}.
\end{equation}
which indicates that $\nabla\rho\times\nabla P_{\rm tot}$ is the source that drives the gas vorticity.
%where $P_{\rm tot}=P+P_{\rm CR}$. Figure~\ref{fig:baroclinic} shows the baroclinic source term at $t=1.7$ Myr.
Compared with the CR-poor model, the CR-rich model exhibits a more extended
baroclinic structure around the clump and in the downstream region.
%This reflects the additional vorticity source associated with the misalignment between $\nabla\rho$ and $\nabla P_{\rm CR}$.

%%%%%%%%%%%%%%%%%%%%%%% energy exchange %%%%%%%%%%%%%%%%%%%%%%%%
\begin{figure}
    \centering
    \includegraphics[width=1.0\linewidth]{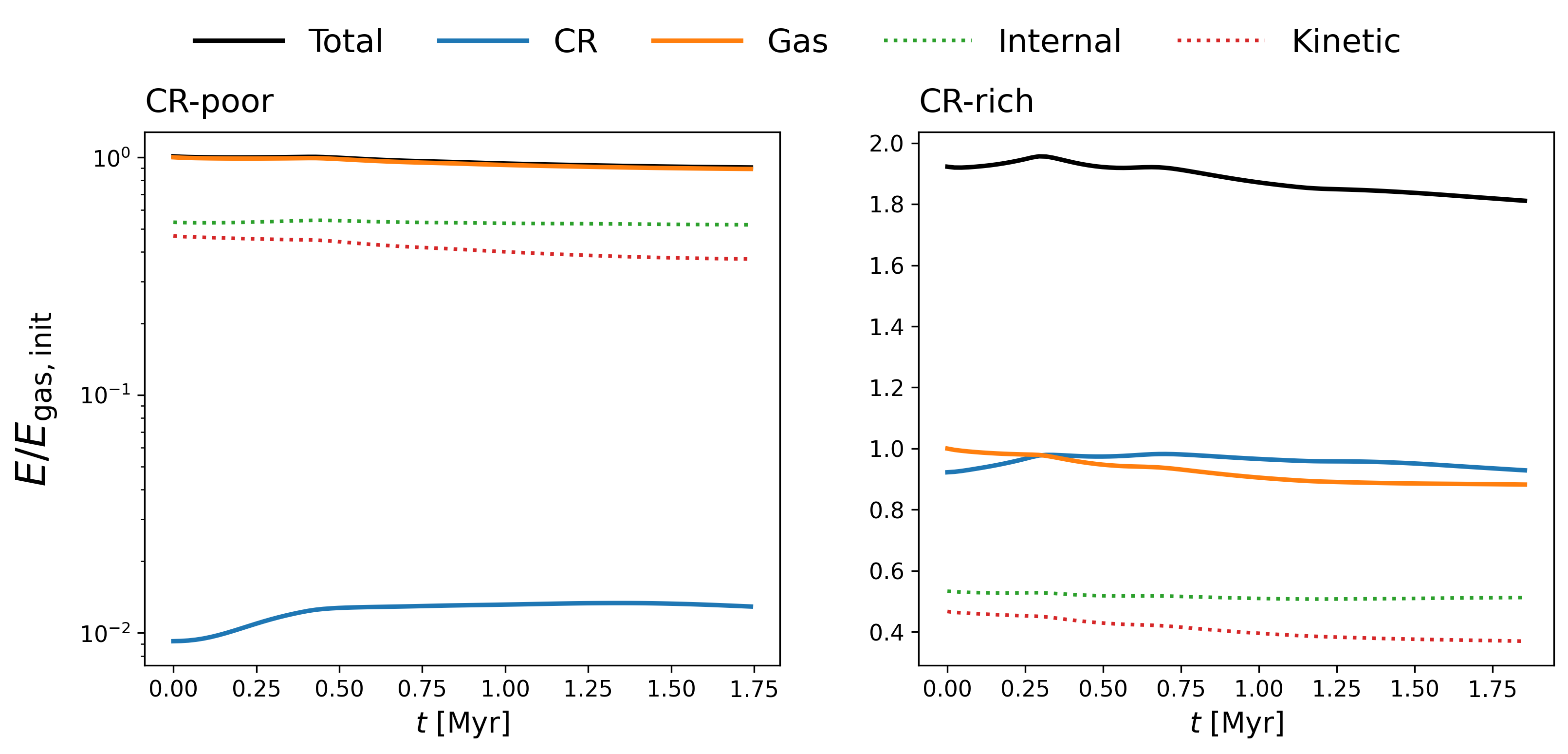}
    \caption{
    Time evolution of the volume-integrated gas and
    CR energies in the CR-poor (left) and CR-rich
    (right) models, evaluated within the cylindrical
    region (see text). %$0\leq R\leq150$ pc and     $0\leq z\leq600$ pc.
    All the energies are normalized by the initial
    total gas energy within the same region.
    The solid curves denote $E_{\rm tot}$,
    $E_{\rm CR}$, and $E_{\rm gas}$,
    while the dotted curves denote $E_{\rm i,gas}$
    and $E_{\rm k,gas}$.
    The left and right panels use logarithmic
    and linear vertical scales, respectively.
    }
    \label{fig:energy_change}
\end{figure}

%We next examine the energy evolution during the clump--wind interaction.
Figure~\ref{fig:energy_change}
shows the time evolution of the volume-integrated
gas and CR energies ($E_{\rm gas}$ and $E_{\rm CR}$, respectively) within the cylindrical region
$0\leq R\leq150$ pc and $0\leq z\leq600$ pc.
The gas energy is decomposed into the internal and
kinetic components (corresponding to the first and second terms in the right-hand side of equation (\ref{eq:gas-ene-den}), respectively),
$E_{\rm gas}=E_{\rm i,gas}+E_{\rm k,gas}$,
and the total energy is defined as
$E_{\rm tot}=E_{\rm gas}+E_{\rm CR}$.
%All energies are normalized by the initial gas energy within the same region.
Note that this is not an energetically closed region, as there are gas inflow and outflow, and radiative cooling and heating.

A fraction of the gas is accelerated beyond its initial velocity, as shown in Figure \ref{fig:1devolution_thin_cr} and \ref{fig:1devolution_thick_cr}. In the CR-rich case, we have seen the enhancement of the turbulence behind the clump and wind acceleration in the outer region.
Nevertheless, Figure~\ref{fig:energy_change} shows that the gas does work on CRs in the vast majority of the region, causing an increase of $E_{\rm CR}$ in both models.
%the CR energy increases during the clump--wind interaction. Although CR pressure promotes the development of rotational gas motions in the downstream wake of the CR-rich model, 
The energy gain associated
with gas compression near the shock is expected
to dominate over the energy transferred from
CRs to the gas.
%The evolution of the volume-integrated energies also reflects energy transport across the boundaries of the integration region.

%%%%%%%%%%%%%%%%%%%%%%%% CR spectrum %%%%%%%%%%%%%%%%%%%%%%%%%%%

\subsection{CR Spectrum}
\label{subsec:cr_acceleration}

\begin{figure}
    \centering
    \includegraphics[width=1.0\linewidth]{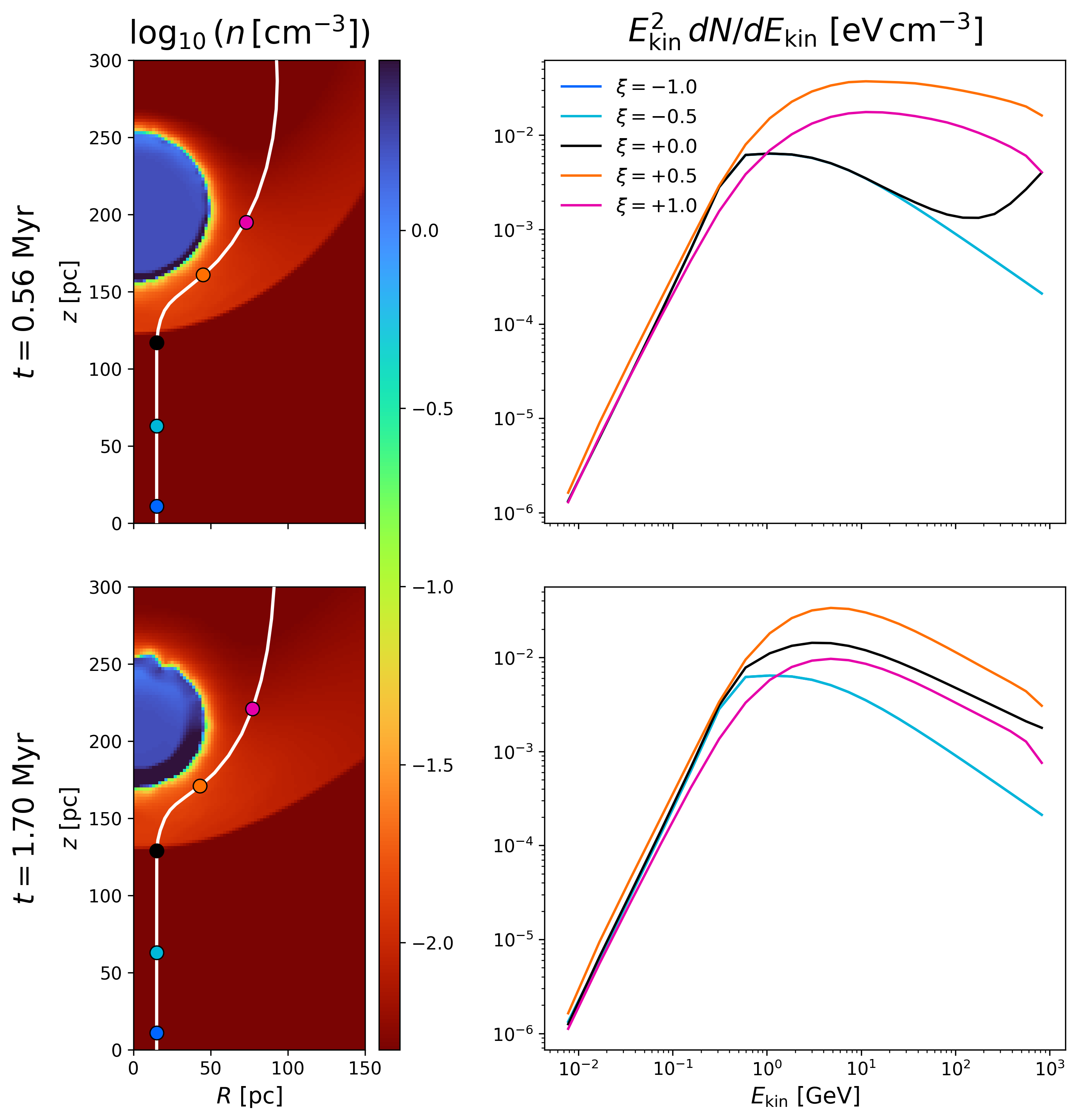}
    \caption{
    Local CR proton spectra along the representative streamlines in the
    CR-poor model.
    The top and bottom rows correspond to $t=0.56$ and $1.7$ Myr,
    respectively.
    The left panels show the gas density distribution, $n$,
    together with the representative streamline (white line) and the
    positions at which the CR spectra are plotted (colored filled circles).
    %The middle panels show the CR momentum distribution, $dn_{\rm CR}/dp$, and 
    The right panels show the 
    spectra, $E_{\rm kin}^{2}dN/dE_{\rm kin}$, with the same color as the corresponding circle in the left panel.
    %The spectra and corresponding crosses are color-coded by the shock-relative position, $\xi=(s-s_{\rm sh})/l_{\rm diff}$.
    }
    \label{fig:crspec_poor}
\end{figure}

Gas compression can accelerate CRs and modify their energy spectrum. To plot the local CR spectra as a function of the proton kinetic energy, $E_{\rm kin} \equiv E_{\rm p}-m_{\rm p} c^2$, in Figure \ref{fig:crspec_poor}, we extract the positions along the representative streamlines shown in the left panels. %Figures~\ref{fig:1devolution_thin_cr} and \ref{fig:1devolution_thick_cr}.
For the CR-poor model, we characterize the sampling position relative to the shock front by
\begin{equation}
    \xi \equiv \frac{s-s_{\rm sh}}{l_{\rm diff}},
    \qquad
    l_{\rm diff} = \frac{D(1\,{\rm TeV}/c)}{v_{\rm w}},
\end{equation}
where $s$ is the distance along the streamline and
$s=s_{\rm sh}$ is the shock position.
For our parameter sets,
$l_{\rm diff}\simeq100$ pc.

Figure~\ref{fig:crspec_poor} shows that particles begin to be accelerated at the shock front and that the spectra harden downstream. As shown in Figure \ref{fig:1devolution_thin_cr}, the gas compression continues downstream, which further accelerates CRs.
CR acceleration occurs throughout
the broad compression region rather than being concentrated at a localized shock front.
The gas at $\xi=1.0$ expands and accelerates so that the CR density turns to decline. At 0.56 Myr (upper panel), the CR spectra show irregular behavior. This may be exhibiting transient behavior as the CR distribution approaches a quasi-steady state.
The high-energy part ($>100$ GeV) of the spectrum at $\xi=0$ (black), just outside the shock front, is affected by diffusion from the downstream, while the low-energy part remains unchanged from the initial one.
At 1.7 Myr (lower panel), though the CR spectra become softer than those at 0.56 Myr, the spectral peak energy is significantly larger than the initial value.
%As the diffusion coefficient increases with particle momentum, high-energy CRs can diffuse farther from the shock compression region. The spatial extent of the modified high-energy component is comparable to the characteristic diffusion--advection scale $D(1\,{\rm TeV}/c)/v_{\rm w}\sim100$ pc, consistent with transport of CRs accelerated around the bow shock.

%the strongest spectral modification in the CR-poor model occurs near the bow shock. Both the normalization of the CR spectrum and its high-energy component change rapidly across the shock-relative sampling positions. The modification is not restricted to the immediate shock vicinity: a high-energy excess also extends away from the main compression region.

\begin{figure}
    \centering
    \includegraphics[width=1.0\linewidth]{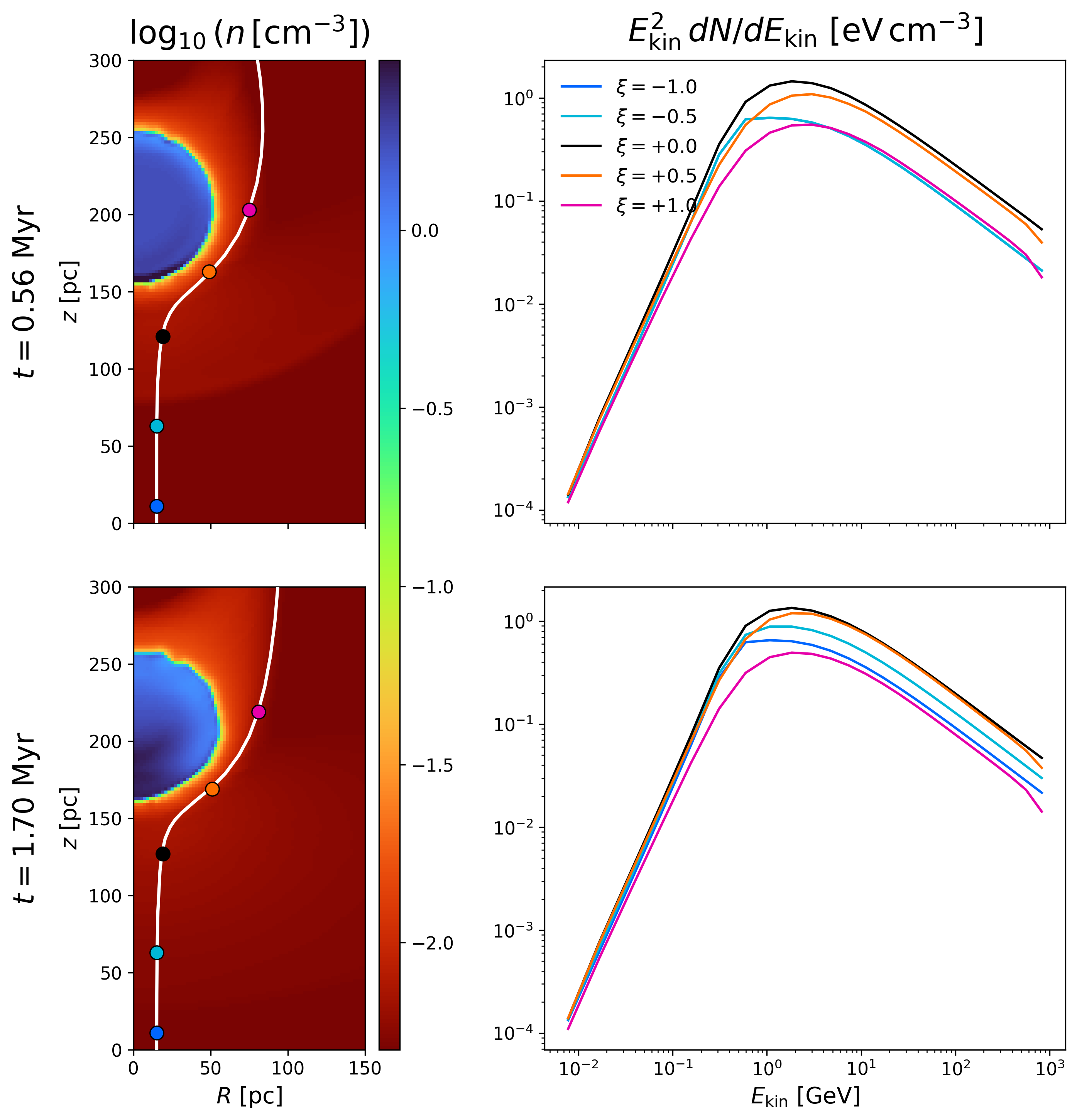}
    \caption{
    Same as Fig.~\ref{fig:crspec_poor}, but for the CR-rich model.
    %Because no comparably well-defined bow shock is present in the CR-rich model, the shock position $s_{\rm sh}$ measured in the CR-poor model is used as a common reference when defining $\xi$.
    }
    \label{fig:crspec_rich}
\end{figure}

In the CR-rich model, as the shock front is not well-defined,
we adopt the same interval between the sampling positions.
As shown in Figure~\ref{fig:crspec_rich}, %shows the corresponding evolution 
%A comparably well-defined bow shock is not present in this model. We therefore use the shock position measured in the CR-poor model only as a common spatial reference when defining $\xi$, which allows the two models to be compared directly.
The spectral evolution is modest
compared with the CR-poor case.
This is because the gas compression behind the shock front is weaker than in the CR-poor case.
%the spectral variation between adjacent sampling positions is smaller and extends over a broader spatial range. This behavior is consistent with 

%%%%%%%%%%%%%%% gamma ray %%%%%%%%%%%%%%%%%%%%%%%
\subsection{Gamma-Ray Emission}

%In our model, 
Gamma rays are produced through inelastic
$pp$ collisions between CR protons and ambient gas in
the Galactic halo.
As shown in Figures~\ref{fig:crspec_poor} and \ref{fig:crspec_rich}, CRs are accelerated
by gas compression in the interaction region between
the clump and the %Galactic wind.
shock front.
The accelerated CRs %accumulate between the bow shock and the clump and 
subsequently diffuse into the
%dense gas
dense clump, while rarefaction waves allow CRs initially
contained in the clump to escape backward.
The penetration of accelerated CRs into the dense
clump enhances the hadronic gamma-ray emission.

To obtain the %observable 
gamma-ray intensity for an observer,
we reconstruct the three-dimensional axisymmetric
distribution from the two-dimensional $(R,z)$
simulation and integrate the emissivity along
the line of sight $dl$ for a side-on view.
The differential gamma-ray intensity is given by
\begin{equation}
    I_\gamma(E_\gamma)
    =
    \frac{1}{4\pi}
    \int_{\rm los}
    q_\gamma(E_\gamma,R,z)\,dl .
\end{equation}
We calculate the energy-integrated intensity over
$1$--$10$ GeV as
\begin{equation}
    I_{1-10\,{\rm GeV}}
    =
    \int_{1\,{\rm GeV}}^{10\,{\rm GeV}}
    %E_\gamma
    I_\gamma(E_\gamma)\,dE_\gamma .
\end{equation}
We define a fixed projected region of interest (ROI)
of $200\times200$ pc centered on the clump.
The same ROI is used at all simulation times and
for both CR models.

\begin{figure}[t]
    \centering
    \includegraphics[width=1.0\linewidth]{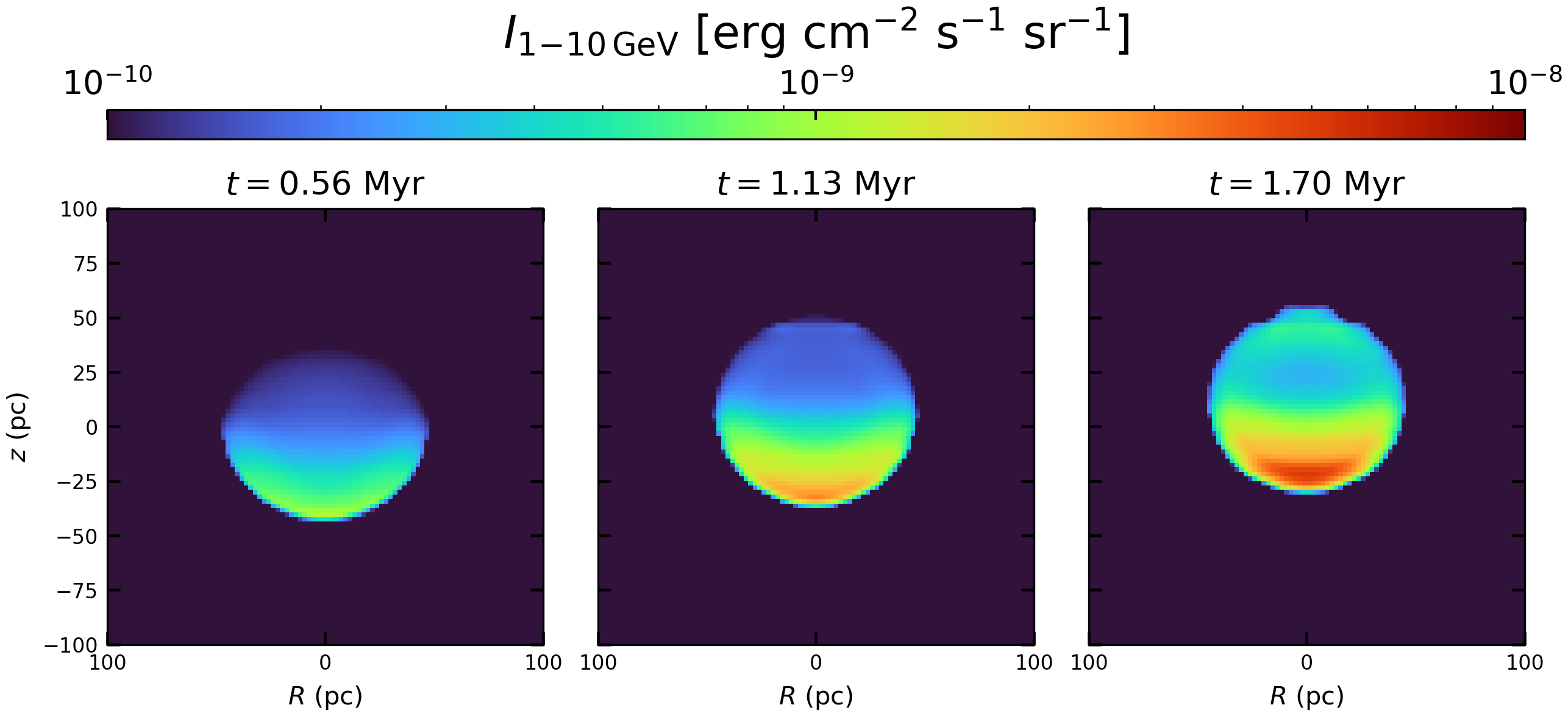}
    \caption{
    Time evolution of the gamma-ray intensity in the CR-poor model,
    integrated over 1--10 GeV.
    The three panels correspond to $t=0.565$, $1.13$, and $1.70$ Myr, from left to right. The image area shown here corresponds to the ROI (see text).
    }
    \label{fig:gamma_intensity_poor}
\end{figure}

\begin{figure}[t]
    \centering
    \includegraphics[width=1.0\linewidth]{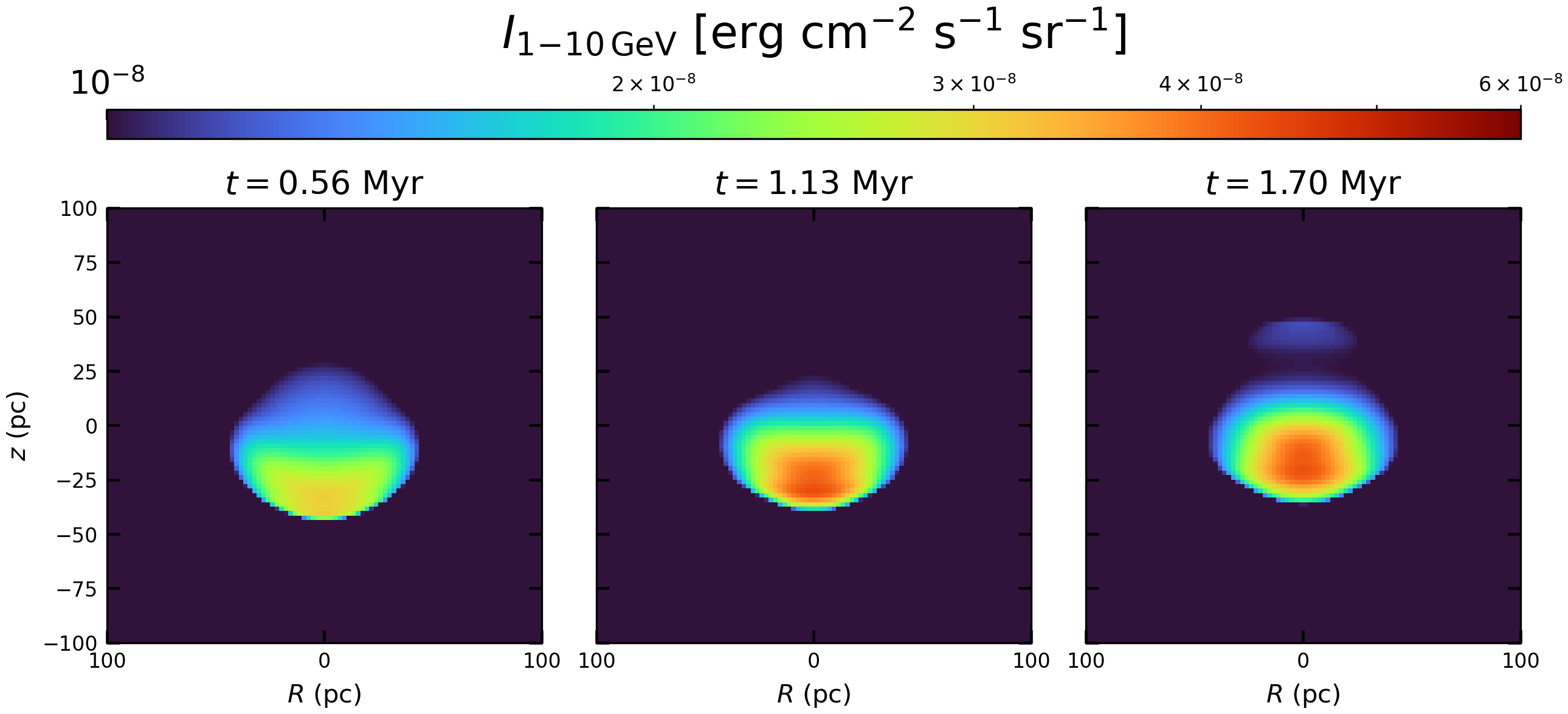}
    \caption{
    Same as Fig.~\ref{fig:gamma_intensity_poor},
    but for the CR-rich model.
    }
    \label{fig:gamma_intensity_rich}
\end{figure}

Figures~\ref{fig:gamma_intensity_poor} and
\ref{fig:gamma_intensity_rich} show the time evolution
of the projected gamma-ray intensity $I_{1-10\,{\rm GeV}}$.
%integrated over $1$--$10$ GeV.
In both models, the emission is concentrated around
the dense clump and becomes brighter with time.
%The emission remains approximately symmetric around the clump, consistent with the diffusive penetration of accelerated CRs into the dense gas.
The CR-rich model naturally produces a higher overall intensity than in the CR-poor model. 
However, even with an initial $P_{\rm CR}$ of only 1\% relative to the CR-rich model, the CR-poor model yields $\sim 10$ \% of its intensity.
%also exhibits a substantial enhancement despite its much lower initial CR density.

In our two models, the gamma-ray intensity reaches
around $10^{-9}$--$10^{-8}\
{\rm erg\,cm^{-2}\,s^{-1}\,sr^{-1}}$
at the bright regions in the clump,
approximately an order of magnitude higher than
the observed surface brightness of the Fermi bubbles,
$\sim 10^{-10}$--$10^{-9}\
{\rm erg\,cm^{-2}\,s^{-1}\,sr^{-1}}$
in the same energy band
\citep{Ackermann2014}.
This suggests that an ensemble of such clumps
distributed %throughout the Galactic wind could
above the Galactic center
account for the observed gamma-ray emission
with a projected filling factor of order $f\sim0.1$. Note that the gamma-ray intensity is not saturated yet, especially for the CR-poor case. We can expect brighter intensities from this kind of clumps.
%The superposition of emission from individual clumps, combined with the finite angular resolution of Fermi-LAT, could produce the extended, diffuse gamma-ray emission observed in the Fermi bubbles.

\begin{figure}[t]
    \centering
    \includegraphics[width=1.0\linewidth]{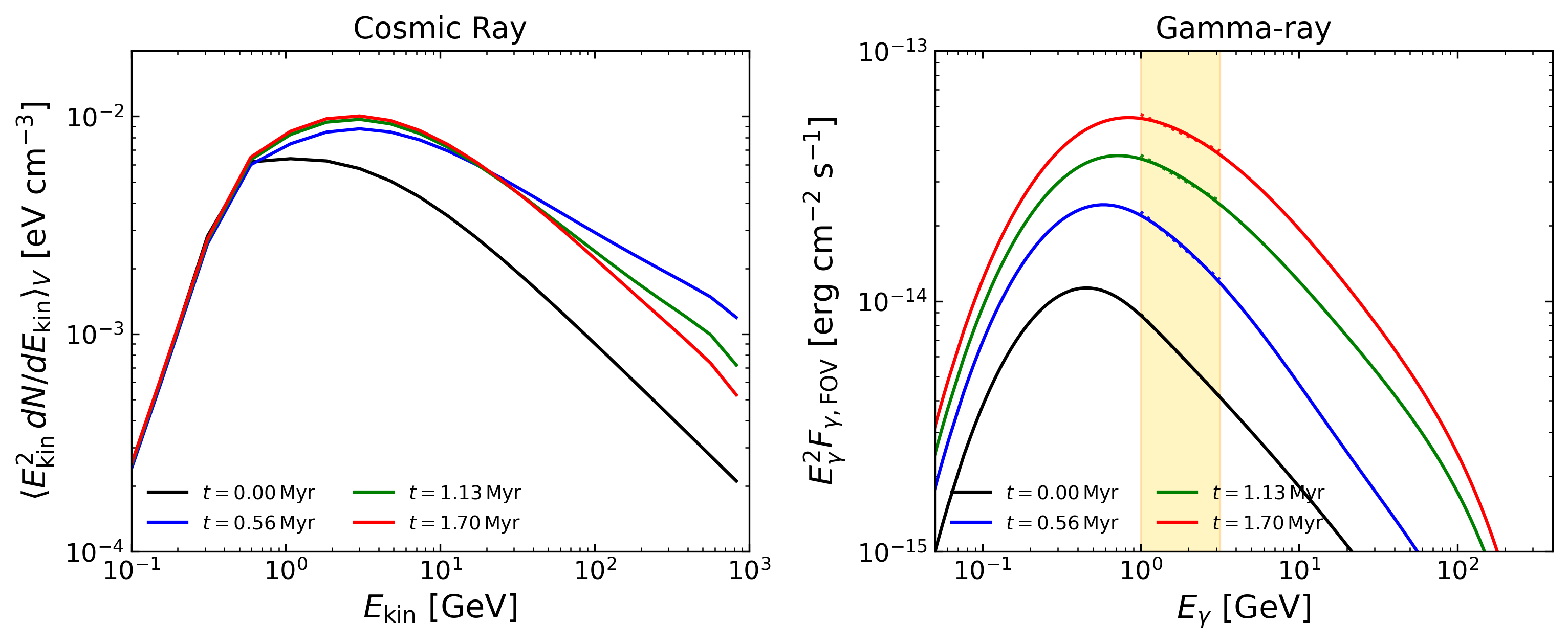}
    \caption{
    Evolution of the CR proton and gamma-ray spectra in the CR-poor model.
    Left: volume-averaged CR %kinetic-energy 
    spectra within the
    three-dimensional region corresponding to the projected ROI
    shown in Figs.~\ref{fig:gamma_intensity_poor} and \ref{fig:gamma_intensity_rich}.
    Right: gamma-ray %flux
    spectra integrated over the same projected ROI for an observer at a distance of $8.5$ kpc from the clump.
    The shaded region indicates the 1--3.16 GeV range used to
    evaluate the photon index in Fig.~\ref{fig:gamma_index_evolution}.
    }
    \label{fig:roi_spectra_poor}
\end{figure}

\begin{figure}[t]
    \centering
    \includegraphics[width=1.0\linewidth]{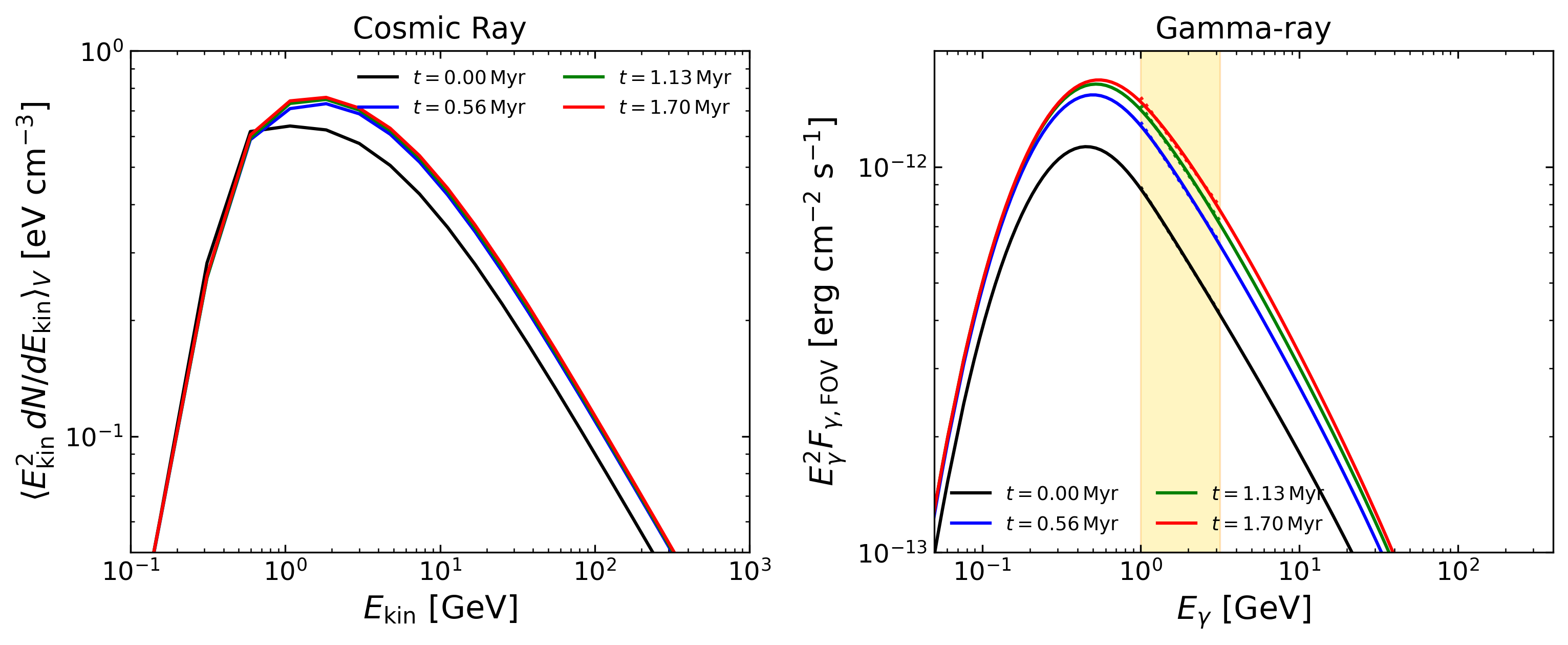}
    \caption{
    Same as Fig.~\ref{fig:roi_spectra_poor},
    but for the CR-rich model.
    }
    \label{fig:roi_spectra_rich}
\end{figure}

%To examine the spectral evolution, we calculate
In Figures~\ref{fig:roi_spectra_poor} and \ref{fig:roi_spectra_rich}, we show the volume-averaged CR proton spectra over the
three-dimensional region that projects into the
fixed ROI, together with the gamma-ray fluxes
integrated over the same projected ROI.
%Figures~\ref{fig:roi_spectra_poor} and \ref{fig:roi_spectra_rich} show the resulting spectra for the CR-poor and CR-rich models, respectively.
In the CR-poor model, the average CR spectrum in the ROI becomes
progressively harder with time, accompanied by
a corresponding hardening. %of the gamma-ray spectrum.
In contrast, the CR-rich model exhibits 
a modest change 
in the spectral shape, despite its
higher gamma-ray flux.
%This difference reflects the stronger localized gas compression in the CR-poor model and the weaker, broader compression caused by CR feedback in the CR-rich model.

\begin{figure}[t]
    \centering
    \includegraphics[width=1.0\linewidth]{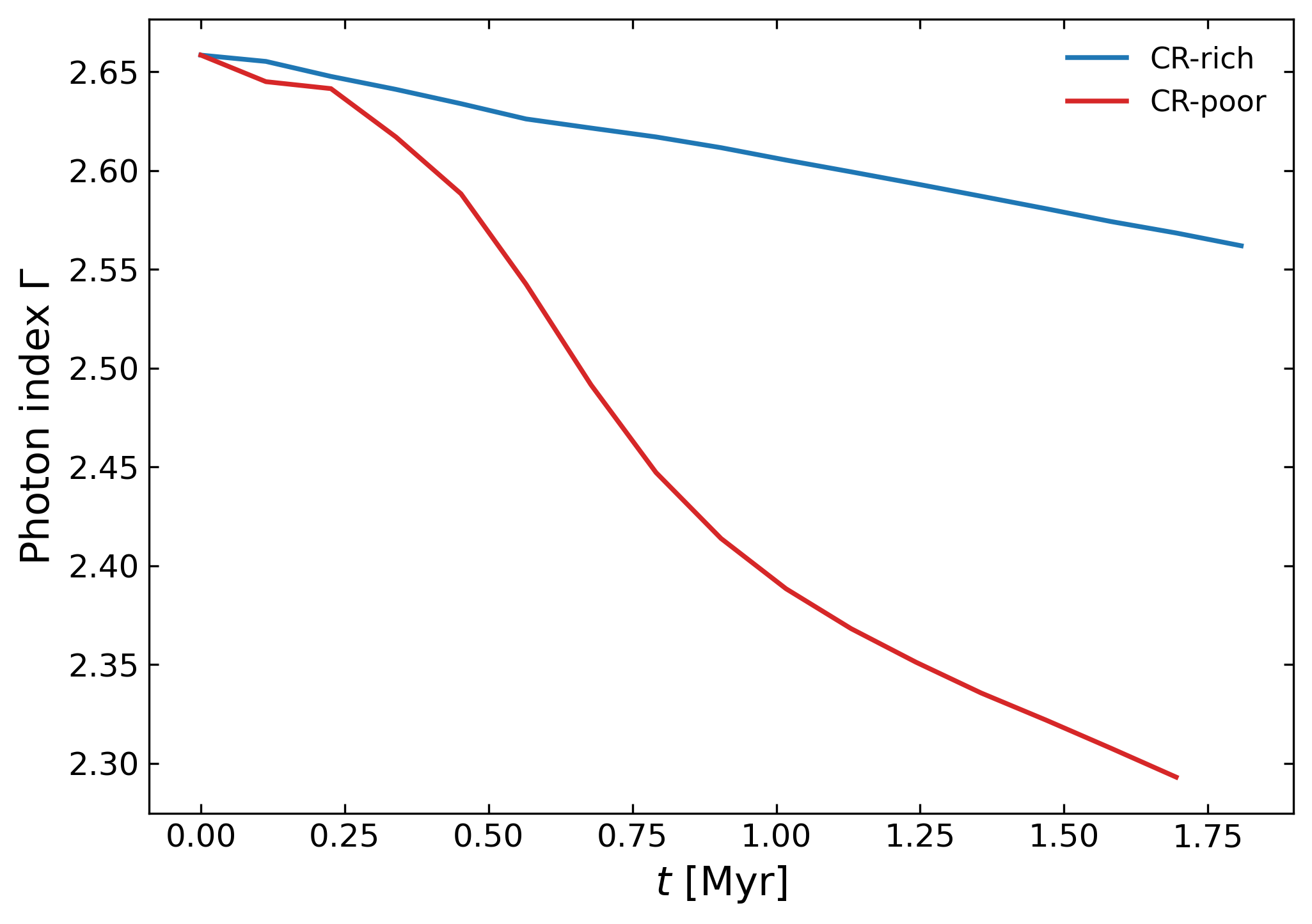}
    \caption{
    Evolution of the gamma-ray photon index measured
    in the $1$--$3.16$ GeV band.
    The red and blue lines show the CR-poor and
    CR-rich models, respectively.
    }
    \label{fig:gamma_index_evolution}
\end{figure}

To quantify the spectral hardening, we fit the
gamma-ray spectra in the $1$--$3.16$ GeV range
with a power law,
$dN_\gamma/dE_\gamma\propto E_\gamma^{-\Gamma}$.
Figure~\ref{fig:gamma_index_evolution} shows the
time evolution of the photon index $\Gamma$.
In the CR-poor model, the photon index decreases
from $\Gamma\simeq2.66$ initially to
$\Gamma\simeq2.29$ at the end of the calculation.
This hard spectral index is consistent with the observed index in the Fermi bubbles. On the other hand,
%whereas 
the CR-rich model shows a smaller change,
from $\Gamma\simeq2.66$ to $\Gamma\simeq2.57$.
The difference in the final index is due to the different gas compression degree in the two models.
%The stronger spectral hardening in the CR-poor model is consistent with efficient CR acceleration at the localized bow shock, while CR feedback in the CR-rich model weakens the compression and reduces the resulting spectral modification.

\section{Discussion}
\label{sec:discussion}

\subsection{Fermi Bubbles}

The Galactic halo is a complex system in which outflowing high-temperature galactic winds, cold accretion flows, and cosmic rays diffusing from the Galactic disk interact.
As the two simulations we have demonstrated in this paper use a limited set of parameters, we are still at a preliminary stage in terms of uncovering the gas and CR interactions in the Galactic halo. However, our results indicate that CRs have a significant impact on the galactic wind and accretion flow, and that accreting gas clumps are a promising candidate for the source of the gamma-ray emission from the Fermi bubbles.
Especially, hard spectral indices we obtained are encouraging for the Fermi bubbles.

In reality, the heating efficiency and diffusion coefficient depend on the position and environment of the clumps, and their velocities and sizes may be distributed over a wide range. Each clump reaccelerates CRs at a different efficiency. The duration of the clump--wind interaction is much longer than our simulation time. While we have verified gamma-rays from CRs that penetrate the clump directly from the acceleration region, the escaped CRs can also collide with gas in other clumps, emitting gamma-rays. The above facts suggest that the global properties of the emission are yet uncertain, but the gamma-ray emission efficiency can be higher than in our simulations.

%The actual interaction, however, depends on the relative motion between the Galactic wind and cold gas clouds. In this study, we assume a stationary clump and a constant wind velocity of $v_w=300\,\mathrm{km\,s^{-1}}$, whereas the wind velocity and cloud motion in the Galactic halo remain uncertain. Clouds entrained in the outflow may have smaller relative velocities, while infalling high-velocity clouds (HVCs) \citep{2007ApJ...670L.113W, 2008ApJ...679L..21L} may encounter the wind at substantially higher velocities. Such variations can modify the strength of gas compression and hence the efficiency of CR acceleration.

%{
%\color{red}
%The duration of the wind--clump interaction is another important uncertainty. Recent hydrodynamical simulations by \citet{Schulreich2026} suggest that radiative cooling and gas condensation can allow infalling HVCs to survive for extended periods in the Galactic halo. If the interaction persists for $\sim10\,\mathrm{Myr}$ or longer, CRs may undergo acceleration over a longer period than considered in our simulations. In particular, infalling clouds could provide favorable conditions for further CR acceleration owing to their larger relative velocities. The resulting spectral evolution would depend on the competition between acceleration and diffusive escape, as well as the dynamical feedback of CRs.
%}

If a projected filling factor of clumps is order $f\sim0.1$, the average gamma-ray intensity can be comparable to the Fermi bubbles' intensity. Recent H I observations have identified hundreds of neutral gas clouds in the Galactic nuclear wind \citep{Lockman2026}.
The clump mass in our simulation is $2.3 \times 10^4 M_\odot$, which is much larger than observed clumps \citep[typically $\sim 10^3 M_\odot$,][]{noon2026}. To consider typical cases in the Fermi bubble, let us halve the clump radius, leading the mass to $\sim 10^3 M_\odot$. The intensity would be reduced by only half, as the CR density would be almost the same. If the descent velocity of the clump to the Galactic disk is $\sim 300~\mbox{km}~\mbox{s}^{-1}$, the mass supply timescale by this clump with a diameter of $50$ pc is $\sim 0.2$ Myr. Thus, the contribution of such a clump to the gas accretion rate is $\sim 2\times 10^{-2}~M_\odot~\mbox{yr}^{-1}$. To maintain star formation in the Galactic center ( $\sim 1 M_\odot~\mbox{yr}^{-1}$), $ \sim 50$ clouds are required to exist in a cylinder with the Fermi bubbles' radius of 3.5 kpc and a height of 50 pc. In this case, the projected filling factor is $\sim 0.3$.

Although the above estimate is very rough, scattered small spots of gamma-ray emission regions are advantageous for explaining the observed surface brightness distribution. If CRs inside the bubbles emit gamma-rays uniformly, the thicker central region appears brighter, but the actual surface brightness is almost uniform. Considering the distance of 8.5 kpc, a bright region of about 50 pc is roughly the same size as the angular resolution of the Fermi LAT at 3 GeV. The gamma-ray fluxes shown in Figures~\ref{fig:roi_spectra_poor} and \ref{fig:roi_spectra_rich} are below the detection limit for a point source.
Therefore, under the influence of foreground radiation, the gamma-ray spots sparsely distributed throughout the bubbles may not be inconsistent with the analysis that concluded the Fermi bubbles have a nearly uniform surface luminosity.

\subsection{CR Shear Acceleration and Viscosity}

In our simulations, we include only the term of the CR energy change due to compression in the diffusion--convection equation.
%In addition to compression and shock acceleration,
Velocity shear generated around the clump may also provide an additional mechanism for CR acceleration, particularly at high energies \citep{1988ApJ...331L..91E, Rieger2019}. As shown in ~\cref{fig:baroclinic}, CR pressure modifies the baroclinic generation of vorticity, suggesting that CR feedback can influence the development of complex gas flows.
By introducing the contribution of the strain-rate tensor in the equations, we can include shear acceleration and its backreaction on the gas through anisotropic CR pressure. This may therefore modify both CR acceleration and gas evolution \citep{Webb1989}. Although the present results show stronger localized spectral hardening in the CR-poor model, shear acceleration could also enhance the high-energy CR population in the CR-rich model, where compression-driven acceleration is less efficient (\cref{fig:crspec_rich}).
On the other hand, the reaction of CRs on a fluid acts as viscosity. The turbulence induced by CRs in the CR-rich case may be softened by the CR viscosity.
%Such additional acceleration processes may contribute to the hard gamma-ray spectrum of the Fermi bubbles. Future simulations incorporating shear acceleration and its dynamical feedback would help clarify this possibility.

\subsection{Shock Identification}

Detecting signatures of the shocks induced by accreting clumps is essential to verify the gas and CR interactions in the Galactic halo. However, in our cases, a prominent temperature jump does not appear at the shock front. The shock identification by X-ray telescopes or line emissions from metals \citep[e.g.,][]{2014ApJ...791...41H} is difficult.
It might actually be more promising to search for evidence of a temperature drop around clumps.
%Finally, the modification of the bow-shock structure by CR pressure may have implications for thermal X-ray observations. A strong bow shock can produce localized compression and heating of the gas, potentially enhancing its thermal X-ray emission, whereas CR feedback can broaden the compression region and reduce the strength of the localized shock. X-ray observations of the Fermi and eROSITA bubbles provide constraints on the thermal structure of Galactic outflows \citep{Sarkar2017}. Although we do not calculate the X-ray emission in this study, the predicted differences in gas density and temperature may offer an additional way to investigate CR feedback. 
While we have neglected electron CRs, they can be accelerated at the bow shocks. However, radio synchrotron emission from electron CRs around clouds is hard to detect \citep{2016JCAP...11..021L}.
Future simulations including a wider range of cloud velocities, long-term turbulent evolution, and multiwavelength emission may provide a better method to identify the bow shocks. %will help clarify the properties of wind--clump interactions. %to the nonthermal emission observed in the Galactic halo.

\section{Summary}
\label{sec:summary}

We have investigated the interaction between a hot
Galactic wind and a cold, dense gas clump using
two-dimensional axisymmetric hydrodynamic simulations
coupled with CR transport.
We have taken into account the CR pressure on the gas, CR acceleration by gas compression, radiative cooling, photoelectric heating, and thermal conduction in our simulations. We have compared two cases: CR-poor ($P_{\rm CR,0}=0.01\,P_0$) and CR-rich ($P_{\rm CR,0}=P_0$) cases.
Our results demonstrate that CR pressure significantly
modifies the gas compression and flow structures
around the clump.
While a sharp isothermal bow shock is formed in the CR-poor case, CR pressure weakens the shock with a smoother and broader pressure profile in the CR-rich case. In both cases, energy was transferred overall from the gas to CRs, where energy transfer via the shock wave is the primary contributor. However, in a part of the downstream, wind gas is accelerated faster than the initial velocity. In the CR-rich model, the wind acceleration and the turbulence behind the cloud are enhanced by the CR effects.
These differences in the gas motion reflect on
%thereby affecting 
the efficiency
of CR reacceleration.
In the CR-poor model, strong gas compression behind
the bow shock leads to significant CR spectral
hardening.
In contrast, in the CR-rich model, %the CR pressure gradient decelerates the incoming wind and produces a broader and 
the weaker compression structure
suppresses the efficiency of CR reacceleration.

We also estimate hadronic gamma-ray emission from CRs based on our simulations. CRs accelerated behind the bow shock penetrate into the dense cloud via diffusion. Most of the hadronic gamma-rays are emitted from the cloud.
%In both models, reaccelerated CRs penetrate the dense clump and enhance hadronic gamma-ray emission.
The resulting gamma-ray photon index in the
1--3.16 GeV range decreases from
$\Gamma \simeq 2.66$ to $\Gamma \simeq 2.29$
in the CR-poor model, whereas the CR-rich model
shows a more modest hardening to $\Gamma \simeq 2.57$.
The gamma-ray intensity around the clump reaches
$\sim 10^{-8}\,\mathrm{erg\,cm^{-2}\,s^{-1}\,sr^{-1}}$
in the 1--10 GeV band, exceeding the characteristic
surface brightness of the Fermi bubbles.
Even in the CR-poor model, the gamma-ray intensity
reaches approximately 10\% of that in the CR-rich
model, despite an initial CR pressure 100 times smaller.

These results suggest that an ensemble of dense
clumps interacting with the Galactic wind may
contribute to the gamma-ray emission of the
Fermi bubbles.
If the projected filling factor of such clumps
is of order $f\sim 0.1$, their collective emission
could reproduce the observed surface brightness.
Although the global emission properties depend
on the distribution and evolution of clumps,
our findings demonstrate the importance of
self-consistently treating CR transport and
gas dynamics in a multiphase Galactic halo.

\begin{acknowledgments}

This work is supported by the joint research program of the Institute for Cosmic Ray Research (ICRR), the University of Tokyo, and KAKENHI grant Nos. 23H04899, 24H00025, and 25K07352 (K.A.), 24K00677, and 25H00394 (J.S.).

\end{acknowledgments}

\bibliography{sample701}{}
\bibliographystyle{aasjournalv7}

%% The collaboration and author truncation commands are used.  It has to


\end{document}